\documentclass[a4paper,11pt]{article}
\pdfoutput=1 

\usepackage{jheppub} 

\usepackage[T1]{fontenc} 

\usepackage{graphicx}
\usepackage{amsmath,amssymb,amsfonts}      
\usepackage{slashed}

\newcommand{\Lg}{\mathcal{L}}
\newcommand{\hn}{h}
\newcommand{\Hn}{{H^0}}
\newcommand{\An}{A^0}
\newcommand{\Hp}{{H^\pm}}
\newcommand{\smo}{\textsc{SModelS}}
\newcommand{\smotwo}{\textsc{SModelS}~v2}
\newcommand{\smoone}{\textsc{SModelS}~v1}
\newcommand{\eg}{\textit{e.g.}}
\newcommand{\ie}{\textit{i.e.}}
\newcommand{\cf}{\textit{cf.}}
\newcommand{\calchep}{\textsc{CalcHep}}
\newcommand{\micromegas}{\textsc{micrOMEGAs}}
\newcommand{\micromegasversion}{\textsc{micrOMEGAs}~v5.2.7.a}

\newcommand{\Ztwo}{\ensuremath{{\mathcal Z}_2}}
\newcommand{\etmiss}{\ensuremath{{E_T^{\rm{miss}}}}}

\title{\boldmath Constraining new physics with SModelS version 2}

\author[a,b]{Ga{\"e}l Alguero,}
\author[c,d]{Jan Heisig,}
\author[e,f,g]{Charanjit Khosa,} 
\author[a]{Sabine Kraml,} 
\author[h]{Suchita Kulkarni,} 
\author[i]{Andre Lessa,}  
\author[e,f]{Humberto Reyes-Gonz\'alez,} 
\author[j,k]{Wolfgang Waltenberger,} 
\author[l]{Alicia Wongel} 

\affiliation[a]{Laboratoire de Physique Subatomique et de Cosmologie (LPSC), Universit\'e
Grenoble-Alpes, \\ CNRS/IN2P3, 53 Avenue des Martyrs, F-38026 Grenoble, France}
\affiliation[b]{LAPTh, Univ. Grenoble Alpes, USMB, CNRS, F-74940 Annecy, France}
\affiliation[c]{Institute for Theoretical Particle Physics and Cosmology, RWTH Aachen University, Sommerfeldstr. 16, D-52056 Aachen, Germany}
\affiliation[d]{Centre for Cosmology, Particle Physics and Phenomenology (CP3), Universit\'e catholique de Louvain, 2 Chemin du Cyclotron, B-1348 Louvain-la-Neuve, Belgium}
\affiliation[e]{Department of Physics, University of Genova, Via Dodecaneso 33, 16146 Genova, Italy}
\affiliation[f]{INFN, Sezione di Genova, Via Dodenasco 33, I-16146 Genova, Italy}
\affiliation[g]{H.H.~Wills Physics Laboratory, University of Bristol, Tyndall Avenue, Bristol BS8 1TL, UK}
\affiliation[h]{Institute of Physics, NAWI Graz, University of Graz, Universit\"atsplatz 5, A-8010 Graz, Austria}
\affiliation[i]{Centro de Ci\^encias Naturais e Humanas, Universidade Federal do ABC, Santo Andr\'e, 09210-580 SP, Brazil}
\affiliation[j]{Institut f\"ur Hochenergiephysik,  \"Osterreichische Akademie der Wissenschaften, Nikolsdorfer Gasse 18, A-1050 Wien, Austria}
\affiliation[k]{University of Vienna, Faculty of Physics, Boltzmanngasse 5, A-1090 Wien, Austria}
\affiliation[l]{DESY, Notkestraße 85, 22607 Hamburg, Germany}

\emailAdd{alguero@lpsc.in2p3.fr}
\emailAdd{heisig@physik.rwth-aachen.de}
\emailAdd{khosacharanjit@gmail.com}
\emailAdd{sabine.kraml@lpsc.in2p3.fr}
\emailAdd{suchita.kulkarni@uni-graz.at}
\emailAdd{andre.lessa@ufabc.edu.br}
\emailAdd{charanjit.kaur@bristol.ac.uk}
\emailAdd{humbertoalonso.reyesgonzlez@edu.unige.it}
\emailAdd{walten@hephy.oeaw.ac.at}
\emailAdd{alicia.wongel@gmail.com}

\abstract{We present version~2 of \smo, a program package for the fast reinterpretation of LHC searches for new physics on the basis of simplified model results. The major novelty of the \smotwo\ series is an extended topology description with a flexible number of particle attributes, 
such as spin, charge, decay width, etc. This enables, in particular, the treatment of a wide range of signatures with long-lived particles.  
Moreover, constraints from prompt and long-lived searches can be evaluated simultaneously in the same run. 
The current database includes results from searches for heavy stable charged particles, disappearing tracks, displaced jets and displaced leptons, in addition to a large number of prompt searches. 
The capabilities of the program are demonstrated by two physics applications: constraints on long-lived charged scalars in the scotogenic model, and constraints on the electroweak-ino sector in the Minimal Supersymmetric Standard Model.}

\begin{document} 
\maketitle
\flushbottom

\section{Introduction}

Run~2 of the Large Hadron Collider (LHC) has presented us with an incredible performance and an exciting physics programme in the quest for new physics beyond the Standard Model (BSM). 
Given the null results so far in the plethora of searches for new particles, it becomes increasingly clear that the effects of BSM physics must either be manifest at higher energy scales, or be much more subtle and/or complicated to find than originally hoped for. Consequently, searches for new physics by the LHC experimental collaborations have intensified and widened in scope. One of the directions in which many new analyses have been directed is searches for new long-lived particles (LLPs), and many interesting new results in the pursuit of both, ``prompt'' and ``long-lived'' new physics have been coming out for the Run~2 data. 

On the phenomenology side, reinterpretation efforts have intensified, with much closer theory-experiment interaction than seemed possible at the beginning of the LHC era \cite{Kraml:2012sg,LHCRiF:2020xtr,Cranmer:2021urp}. It is in this spirit that public software tools for the reinterpretation of the LHC results have been developed, to allow for reinterpretation studies outside the experimental collaborations, help obtain a global and coherent view of how the experimental data constrain BSM scenarios, and guide future experimental and theoretical explorations. 
See Ref.~\cite{LHCRiF:2020xtr} for the status and recommendations after Run~2, as well as an overview of reinterpretation methods and tools. 

For BSM theories with a \Ztwo-like symmetry, the \smo\ package~\cite{Kraml:2013mwa,Ambrogi:2017neo,Dutta:2018ioj,Ambrogi:2018ujg,Khosa:2020zar,Alguero:2020grj,Alguero:2020yhu} provides a particularly efficient way to reuse and reinterpret the results from LHC searches for new particles. 
\smo\ is based on the concept of simplified models; it decomposes the signatures of full BSM scenarios into simplified model components (here called \emph{topologies}) which are then confronted against the experimental constraints from a large database of results. 
Since this does not involve any Monte Carlo event simulation, \smo\ is extremely fast and thus particularly well suited for model surveys, including large scans.  The code and its vast database were also exploited recently in a new statistical learning algorithm that aims at identifying dispersed signals in the slew of LHC results~\cite{Waltenberger:2020ygp}.   

While very powerful, the \smo\ approach to testing BSM theories involves a number of approximations. In particular, in the \smoone\ series, the simplified model description involved only the structure of the topology (number of vertices in each branch, and number and type of SM final states in each vertex) and the masses of the BSM particles.  
Other properties like spin or color representation of the BSM particles, which might influence the kinematic distributions of the final state, were ignored. For many LHC searches with rather inclusive kinematical selection, in particular most searches for R-parity conserving supersymmetry (SUSY), this is a reasonable approximation~\cite{Kraml:2013mwa,Edelhauser:2014ena,Edelhauser:2015ksa,Kraml:2016eti}. Nonetheless, there is need for a more refined simplified model description in \smo, in particular to be able to include the large variety of decay-width (lifetime) dependent LLP results.  

We therefore present in this paper version~2 of \smo.\footnote{The concrete version which this paper is based on is v2.1.} The most important new development in this new version is the introduction of a \emph{particle class}, enabling an extended topology description with a flexible number of attributes for the BSM particles, 
such as spin, charge, decay width, etc. As mentioned above, this allows in particular for a better treatment of LLP signatures.  
Moreover, constraints from prompt and long-lived searches can be evaluated simultaneously in the same run. 
On the database side, \smotwo\ includes results from searches for heavy stable charged particles (HSCPs), disappearing tracks, displaced jets and displaced leptons, in addition to a large number of prompt searches. 

In the following, we describe in detail the technical novelties in \smotwo\ and demonstrate the capabilities of the program by means of two physics applications: the scotogenic model~\cite{Ma:2006km,Kubo:2006yx} with either a scalar or a fermionic dark matter (DM) candidate, and the electroweak (EW)-ino sector of the Minimal Supersymmetric Standard Model (MSSM)~\cite{Haber:1993wf,Baer:1995tb,Drees:1995hj,Nilles:1995ci}. The appendices give further details on the current database, the development of home-grown HSCP efficiency maps, and the interface to  \micromegas~\cite{Belanger:2013oya,Belanger:2018ccd,Barducci:2016pcb}. 
It is assumed that the reader is familiar with the concepts and working principle of \smo\@. If this is not the case, we refer to \cite{Kraml:2013mwa} and the online documentation \cite{smodels:wiki} for a more detailed introduction.

\section{Extension of the topology description in \smotwo}

The most significant novelty in \smotwo\ is the introduction of \emph{particle objects}, which allows for more flexibility when dealing with simplified models. In the \smoone\ series  \cite{Kraml:2013mwa,Ambrogi:2017neo,Ambrogi:2018ujg,Alguero:2020grj}, only the masses of intermediate BSM particles were used for describing the simplified model topologies tested by the database or obtained from the decomposition of the input model. Furthermore all final states, whether they are SM particles or (meta-)stable BSM particles, were described by simple labels (strings). Hence, despite the handling of a large variety of simplified model constraints, \smoone\ was not able to deal with width-dependent results, such as searches for displaced decays, or searches which depend on additional information concerning the BSM particles, such as their spin or color representation.

\subsection{Particle class} \label{sec:particleClass}

\begin{figure}\centering
	\hspace*{4mm}\includegraphics[scale=0.07]{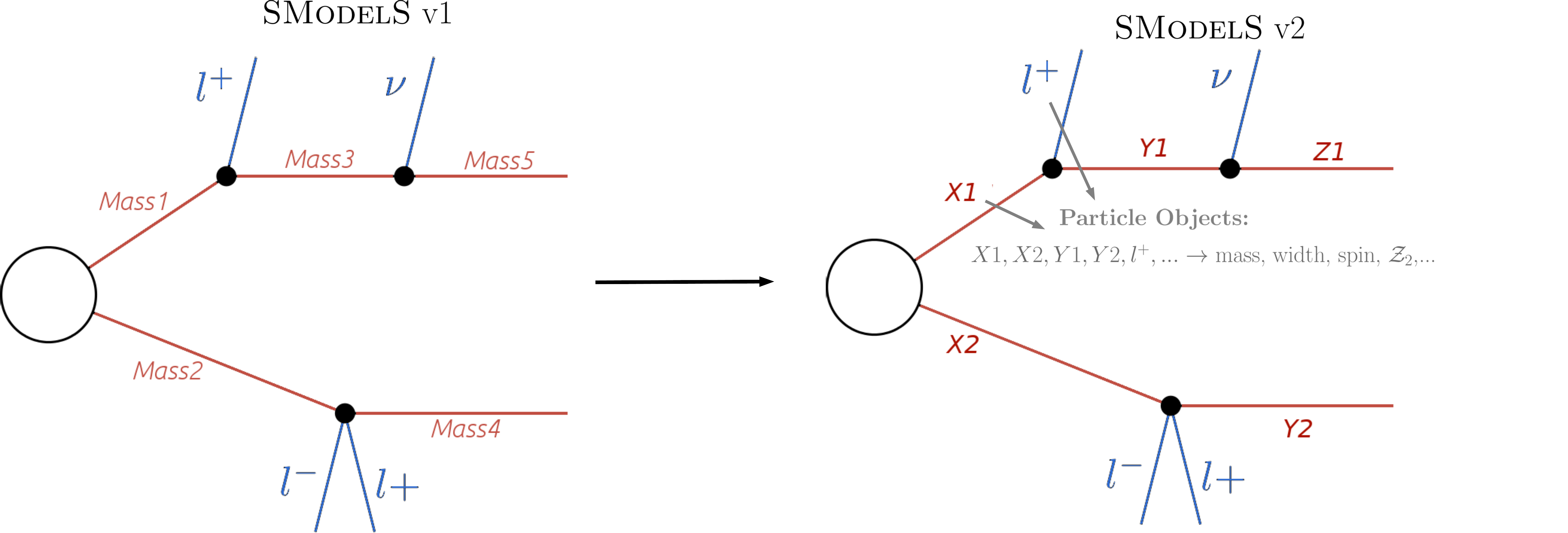}
\caption{Comparison between the simplified model description in \smoone\ (left) and in \smotwo\ (right). For the latter, the newly introduced particle class is used to describe all particles appearing in the topology.} \label{fig:elcomp}
\end{figure}

In the \smotwo\ series, the introduction of a \emph{particle class} replaces the simple list of particle masses by full objects, which can carry any desirable number of properties, such as mass, width, spin, electric charge, etc. As illustrated in Fig.~\ref{fig:elcomp}, these objects are henceforth the fundamental building blocks for describing the experimental results in the database, and for decomposing the input model. When matching the signal topologies of the input model onto those in the database, the comparison is made at the level of particle objects; particle objects with the same properties are considered as equal, independent of their labels.  

At present, the following attributes (properties) are considered: 
\begin{itemize}
	\item \texttt{Z2parity}: the $\mathcal{Z}_2$-type parity of the particle ($-1$ for odd particles and $+1$ for even particles).
	\item \texttt{spin}: the particle spin ($1/2$ for fermions, $0$ for scalars, ...)
	\item \texttt{colordim}: the color representation ($1$ for singlets, $3$ for triplets and $8$ for octets)
	\item \texttt{eCharge}: the electric charge ($-1$ for electrons, $0$ for neutrinos, ...)
	\item \texttt{mass}: the particle mass
	\item \texttt{totalwidth}: the particle total decay width.
\end{itemize}
Note however that the number of attributes is a priori not fixed;  
new ones may be introduced by the user if needed.

In the database of experimental results, if a specific property has no assigned value, it is assumed to be ``arbitrary''. For instance, if a search is sufficiently inclusive to be, to a good approximation, insensitive to the spins of the BSM particles, the spins are left unspecified in the database entry. It is then understood that the corresponding result applies to particles of arbitrary spin.\footnote{This is a good approximation for many \etmiss\ searches, apart from mono-X searches, analyses relying on ISR jets, and analyses explicitly using shape information.}

\subsection{Model input}\label{sec:model}

\begin{figure}\centering
	\includegraphics[scale=0.15]{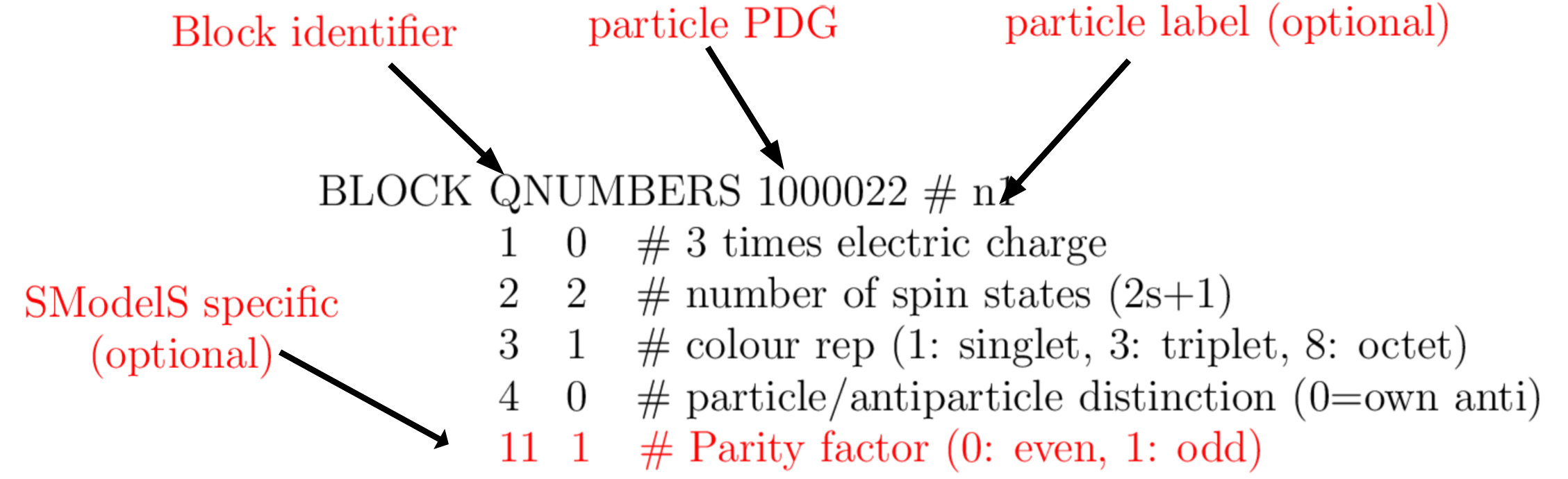}
	\caption{Format for the QNUMBERS block, which can be used to define new particles in \smotwo.} \label{fig:qnumbers}
\end{figure}

The input provided by the user is conveniently split into two files: one containing the definition of BSM particles and their properties (model specification) and one defining their masses, widths, branching ratios and production cross sections (model parameters). The list of BSM particles can be defined in two distinct ways: either by writing a simple file using Python syntax and instantiating the particle objects:\footnote{A few specific model definitions, such as the MSSM, NMSSM and the Inert Doublet Model (IDM), are shipped with the code in the {\tt smodels/share/models} folder and can be used as examples for defining new models.}

\begin{verbatim*}
X=Particle(Z2parity=-1,label='X',pdg=5000021,eCharge=0,colordim=8,spin=1./2)
Y=Particle(Z2parity=-1,label='Y',pdg=5000022,eCharge=0,colordim=1,spin=1./2)  
...
\end{verbatim*}
or by providing a SLHA-type file with QNUMBERS blocks for each BSM particle, as shown in Fig.~\ref{fig:qnumbers}. Note that 
in the QNUMBERS blocks, whether a BSM particle is even or odd is specified by a symmetry factor $S$, $S=0$ or $1$, in key~11. 
This is related to the $\mathcal{Z}_2$(-like) parity, $P_{\mathcal{Z}_2}$, by $P_{\mathcal{Z}_2} = (-1)^S$. If key~11 is not defined, it will be assumed that $S = 1$ (odd particle). 
We point out that such SLHA files containing QNUMBERS blocks for the model definition are generated automatically by the \micromegas-\smo\ interface~\cite{Barducci:2016pcb}, see also Appendix~\ref{app:MOinterface}. Moreover, they can be generated by tools relying on the UFO format~\cite{Degrande:2011ua}, such as MadGraph~\cite{Alwall:2011uj,Alwall:2014hca}.

As in previous versions, the model parameters (masses, decays, ...) 
can be provided either as an SLHA file containing MASS, DECAY and XSECTION blocks, or an LHE file containing parton level events.
The typical usage is to keep the BSM model fixed and test distinct points of its parameter space. In this case, one needs a single file with the list of BSM particles (set, \eg, in the {\tt parameters.ini} file) plus the distinct parameter files for each point of parameter space, which one can also conveniently loop over.
Details and explicit examples are given in the online documentation~\cite{smodels:wiki}.

\subsection{Decomposition for non-prompt decays} \label{sec:decomp}

\begin{figure}
	\begin{center}
		\includegraphics[scale=0.1]{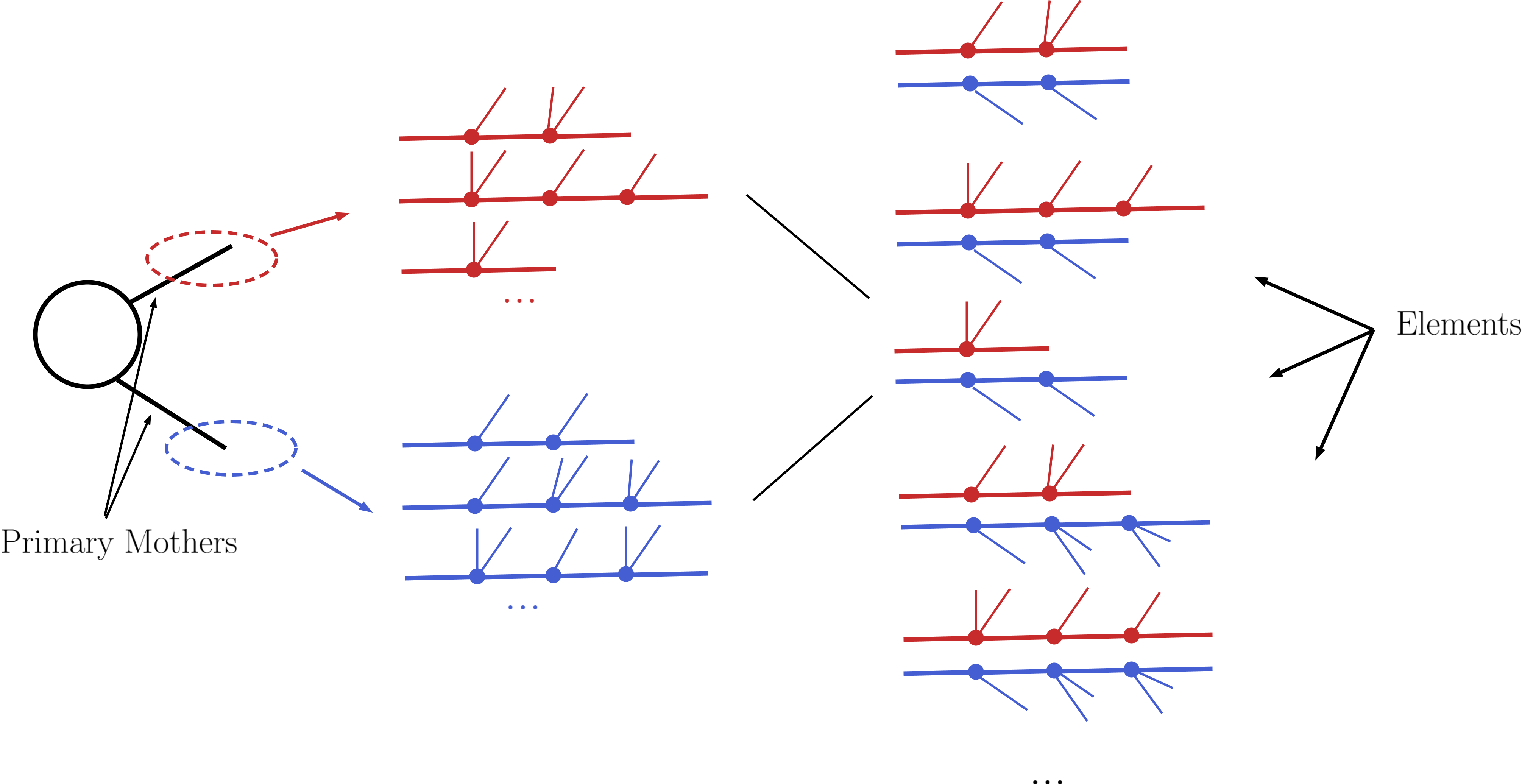}
	\end{center}
	\caption{Schematic representation of how a full model is decomposed into simplified model topologies within \smo.  The input production cross sections are used to defined the primary mothers, which originate a list of possible branches for all the corresponding decay channels of the mother. The branches are then combined in pairs (according to the primary mothers appearing in the production cross sections) to form simplified model topologies, here called elements.} \label{fig:decomp}
\end{figure}

In \smotwo\ the general procedure for decomposing the full BSM model into simplified model topologies is similar to the one in previous versions: the decomposition makes use of the production cross sections to identify the primary mothers produced in the hard scattering processes, which are then (cascade-)decayed according to the decay channels given by the input model. This procedure is followed until all unstable BSM daughters have decayed, as illustrated in Fig.~\ref{fig:decomp}. 
All possible cascade decays of each primary mother are then combined to generate a list of simplified model topologies (also called \emph{elements}), which approximates the input model by a coherent sum of elements. 

\begin{figure}
	\begin{center}
		\includegraphics[scale=0.12]{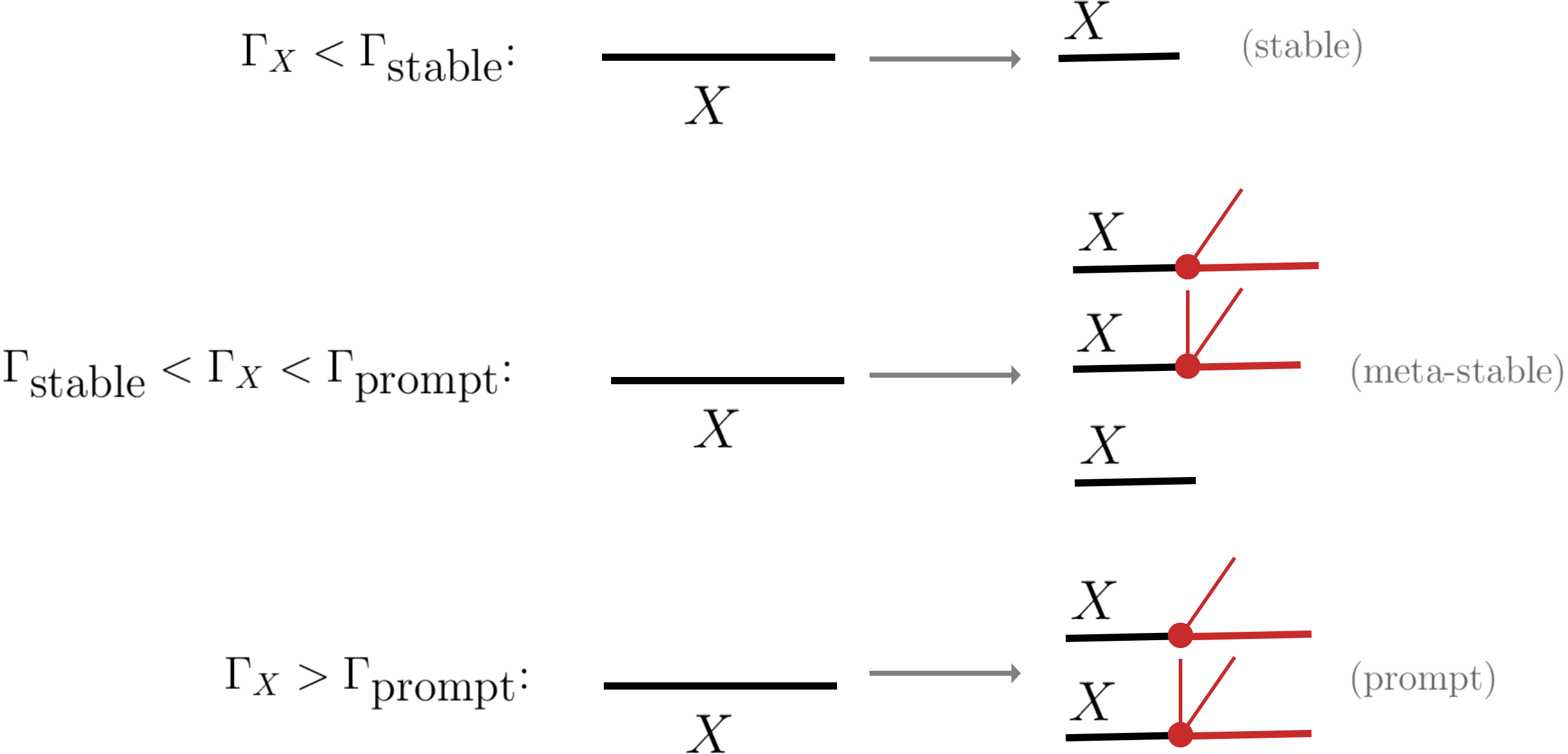}
	\end{center}
	\caption{Illustration of how the decays of a given BSM particle are handled according to the particle's decay width (see text for details).} \label{fig:decays}
\end{figure}

However, the way non-prompt decays are handled has changed significantly. The behavior is controlled by two parameters: 
\begin{itemize}
	\item \texttt{promptWidth}: minimum width for a particle to be considered decaying promptly
	\item \texttt{stableWidth}: maximum width for a particle to be considered detector stable
\end{itemize}
which can be set in the {\tt parameters.ini} file. The default values are $10^{-8}$~GeV for \texttt{promptWidth} and $10^{-25}$~GeV for \texttt{stableWidth}. 
If the total decay width of a given BSM particle is larger than \texttt{promptWidth}, the particle is considered to decay promptly and will never appear as a final state.\footnote{In the current version, the electric charge, color and spin of promptly decaying BSM particles are ignored, which simplifies the decomposition procedure. This is well justified for the  prompt searches in the current database, which are largely insensitive to these quantum numbers.} On the other hand, if its width is smaller than \texttt{stableWidth}, the particle is considered stable on detector scales and all of its decays are ignored. Finally, if the width lies between \texttt{stableWidth} and \texttt{promptWidth}, all relevant topologies are generated where the particle can appear as an intermediate (decayed) or final-state particle.  These cases are shown in Fig.~\ref{fig:decays}. They allow the decomposition procedure to generate topologies where a meta-stable particle can appear either as an intermediate state or a final state, thus allowing \smotwo\ to simultaneously test the input model against searches for both prompt and displaced decays (see Section~\ref{sec:database} for more details).

\subsection{Results description in the database}\label{sec:database}

In previous \smo\ versions, the database of experimental results was limited to upper limit (UL) or efficiency map (EM) results parametrized as a function of the BSM masses ($M_i$) appearing in the simplified model topology. 
Consequently, only searches for promptly decaying BSM particles, or for BSM particles which are stable on detector scales could be included. In order to describe searches for non-prompt particles decaying inside the detector, in \smo\ the ULs or EMs were extended to include both the mass and the width ($\Gamma_i$) of the BSM particles:
\begin{equation}
    \epsilon(M_i) \to \epsilon(M_i,\Gamma_i) \;\mbox{ and }\; \sigma_\mathrm{UL}(M_i) \to \sigma_\mathrm{UL}(M_i,\Gamma_i),
\end{equation} 
where $\epsilon$ represents an efficiency\footnote{In \smo\ terminology, ``efficiencies''  are values of acceptance $\times$ efficiency, $\epsilon\equiv{\cal A}\varepsilon$.} 
and $\sigma_\mathrm{UL}$ an upper limit on the production cross section.

However, most of the (prompt) results contained in the \smo\ database have no explicit width dependence, since they implicitly assume all decays to be prompt and the last BSM particle in the cascade decay to be stable. 
In this case, if the total width of the decaying particle is smaller than {\tt promptWidth}, it is necessary to rescale the original efficiencies or upper limits in order to take into account the width dependence given the implicit assumption of a prompt decay. 
Within \smotwo\ this is done by reweighting the efficiency or upper limit whenever the width dependence is not explicitly specified by the experimental result. The reweighting corresponds to evaluating the approximate fraction of prompt decays appearing in the input topology and the fraction of decays outside of the detector for the last BSM particle:\footnote{Note that since the efficiency dependence on the width scales as $\epsilon \propto  \mathcal{F}_{\mathrm{prompt}}(\Gamma_i)$, the upper limit on the cross section scales as $\sigma_\mathrm{UL} \propto 1/\epsilon \propto 1/\mathcal{F}_{\mathrm{prompt}}(\Gamma_i)$.}
\begin{equation}
    \epsilon(M_i,\Gamma_i) = \mathcal{F}_{\mathrm{prompt}}(\Gamma_i) \times \epsilon(M_i) \; \mbox{ and } \;  \sigma_\mathrm{UL}(M_i,\Gamma_i) = \sigma_\mathrm{UL}(M_i)/\mathcal{F}_{\mathrm{prompt}}(\Gamma_i)\,,  \label{eq:reweight}
\end{equation}
where
\begin{eqnarray}
\mathcal{F}_{\mathrm{prompt}} & = & \left[\prod_{i=1}^{N-1}  \mathcal{F}_{\mathrm{prompt}} (\Gamma_i)\right] \mathcal{F}_{\mathrm{stable}} (\Gamma_N) \nonumber \\
& = &  \left[\prod_{i=1}^{N-1} \left(1 - \exp(-\Gamma_i L_\mathrm{eff}^\mathrm{inner}\right)\right] \exp(-\Gamma_N L_\mathrm{eff}^\mathrm{outer})\,.
\label{eq:fprompt}
\end{eqnarray}
The term in brackets corresponds to the probability that the first $N-1$ decays take place sufficiently close to the interaction point to be considered prompt while the $N$th decay takes place outside the detector. The effective inner and outer detector sizes ($L_\mathrm{eff} = L/\langle \gamma \beta \rangle$) are taken to be $L_\mathrm{eff}^\mathrm{inner} = 0.769$~mm and $L_\mathrm{eff}^\mathrm{outer} = 7.0$~m.
We also point out that the rescaling defined in eq.~\eqref{eq:reweight} can be trivially extended to results where the efficiency (or upper limit) dependence on the widths is partially known, \eg, when only the dependence on the width of the last decay is provided.

Finally, recall that in \smotwo, the quantum numbers of the intermediate and final BSM particles can also be specified when describing an experimental search. Therefore, it is possible to restrict the search applicability to more specific scenarios, such as topologies with a particular spin assignment for the BSM particles. For instance, as discussed in Section~\ref{sec:llpresults}, the implementation of the ATLAS disappearing track search contains distinct efficiency maps for pair production of scalar and femionic LLPs.

\subsection{Missing topologies}

The constraints provided by \smo\ are obviously limited
by its database and the available set of simplified model interpretations provided by the experimental collaborations or computed by theory groups. Therefore it is interesting to identify classes of missing simplified models (or missing topologies) which are relevant for a given input model, but are not constrained by the database.
As in previous versions, \smo\ provides as output the list of the simplified models with highest cross sections which were not constrained by the database. The lack of constraints can be due to two main reasons:
\begin{enumerate}
    \item no experimental result has considered the specific simplified model or
    \item the model parameters (masses and/or widths) fall outside the range considered by the experimental result.
\end{enumerate}

In \smotwo, the classification of missing topologies has changed significantly with respect to previous versions, since  inclusion of displaced results in the database makes it more difficult to uniquely classify the unconstrained topologies. The strategy adopted in \smotwo\ is to classify all experimental results into displaced or prompt results\footnote{Prompt results are all those which assumes all decays to be prompt and the last BSM particle to be stable (or decay outside the detector). Signatures with HSCPs, for instance, are classified as prompt, since the HSCP is assumed to decay outside the detector. Displaced results on the other hand require at least one decay to take place inside the detector.}
and to consider the unconstrained topologies according to these types of results. 
They are therefore grouped according to the following {\it coverage groups}:
\begin{itemize}
    \item {\it missing (prompt)}: not covered by any prompt-type results. This group corresponds to all topologies which did not match any of the simplified models constrained by prompt results in the database.
    \item  {\it missing (displaced)}: not covered by any displaced-type results. This group corresponds to all topologies which did not match any of the simplified models constrained by displaced results in the database.
    \item {\it missing (all)}: not covered by any type of result. This group corresponds to all topologies which did not match any of the simplified models considered by the prompt and the displaced results in the database.
    \item {\it outside the grid}: this group corresponds to topologies which are matched by at least one experimental result in the database (prompt or displaced), but their parameters (masses and/or widths) fall outside the ranges considered by the results.
\end{itemize}

In addition, the {\it missing (prompt)} group reweights its topology cross sections by the fraction of prompt decays, as defined in eq.~\eqref{eq:fprompt}. The  {\it missing (displaced)} group, on the other hand, reweights its topologies by the fraction of displaced decays.
Since the grouping defined above is somewhat arbitrary, it is possible for the user to redefine them with a few simple changes in the \smo\ code, as detailed in the online manual~\cite{smodels:wiki}.

\section{Database extension}

In this section, we briefly summarize the most important additions to the database. 
More details are provided in the release notes, which come with the program package (also available \href{https://smodels.readthedocs.io/en/latest/ReleaseUpdate.html}{online}~\cite{smodels:wiki}).  
An overview of all 13~TeV results included in the \smo\ 2.1.0 database is given in Appendix~\ref{app:analysistables}.

\subsection{Results from searches for long-lived particles} \label{sec:llpresults}

\paragraph{Heavy stable charged particles:} 
For HSCPs, we have newly included results from the 13~TeV ATLAS search~\cite{Aaboud:2019trc} with 36~fb$^{-1}$ (ATLAS-SUSY-2016-32). 
On the one hand, we implemented the UL on the direct production of HSCPs and $R$-hadrons. On the other hand, we recasted the analysis employing the prescription provided in the auxiliary information of the publication and generated EM results for the 11 topologies listed in Appendix~\ref{app:RecastHSCP}. They contain directly produced HSCPs, as well as HSCPs arising from a 1- or 2-step decay. We included asymmetric and mixed HSCP/\etmiss\ branches containing up to three mass parameters. For the EMs with up to two mass parameters, we included the explicit width-dependence as a third parameter. Details are given in Appendix~\ref{app:RecastHSCP}.

\paragraph{Disappearing tracks:} 
The \smo\ database now contains efficiency maps for one or two charged tracks from searches for long-lived charginos by  ATLAS~\cite{Aaboud:2017mpt} (ATLAS-SUSY-2016-06, 36~fb$^{-1}$) and CMS~\cite{CMS:2020atg} (CMS-EXO-19-010, 101~fb$^{-1}$). Since these analyses can be very sensitive to the LLP decay length, which depends on the LLP boost and consequently on its spin,
the disappearing track analyses were implemented for specific spin assignments.
For the ATLAS analysis, we use the efficiency maps provided by  \cite{Belyaev:2020wok,belyaev_2020_4288736} for both the fermionic (chargino) and the scalar (charged Higgs) LLP cases.
For the CMS analysis, we use the efficiency maps provided by the collaboration; here only the fermion (chargino) case is available.

\paragraph{Displaced jets:} 
In this category, we have included UL results from the ATLAS search \cite{Aaboud:2017iio}, which targets final states with large $\etmiss$ and at least one high-mass displaced vertex with five or more tracks (ATLAS-SUSY-2016-08, $32.8$~fb$^{-1}$). Moreover, we have included EM results for the CMS search for non-prompt jets \cite{Sirunyan:2019gut} with full Run~2 luminosity (CMS-EXO-19-001, $137$~fb$^{-1}$).
 
\paragraph{Displaced leptons:} 
 Here, we have implemented the results from the ATLAS search \cite{ATLAS:2020wjh} for charged leptons with large impact parameters,   ATLAS-SUSY-2018-14, for full Run~2 luminosity ($139$~fb$^{-1}$). Note that this analysis provides not only EMs but also the statistical model\footnote{See \cite{Cranmer:2021urp} for a detailed discussion of why and how to publish the statistical models.} in pyhf JSON format on HEPData! 

\subsection{Results from conventional (prompt) SUSY searches}

Newly added were UL and EM results for the ATLAS gluino/squark searches in the $1\ell$+jets and $0\ell$+jets final states, 
ATLAS-SUSY-2018-10~\cite{ATLAS:2021twp} and ATLAS-SUSY-2018-22~\cite{ATLAS:2020syg}, and for the $0\ell$ stop search 
ATLAS-SUSY-2018-12~\cite{ATLAS:2020dsf}. 
Likewise, UL results were added for the electroweak-ino search in the $WH+\etmiss$ final state, ATLAS-SUSY-2018-23~\cite{ATLAS:2020qlk} (no EMs available in this case). All these analyses are for 13~TeV and full Run~2 luminosity. 

Moreover, we augmented the previously available UL results from the 13~TeV ATLAS electroweak-ino searches in the $WZ+\etmiss$ final state,  
ATLAS-SUSY-2017-03~\cite{Aaboud:2018sua} and ATLAS-SUSY-2018-06~\cite{Aad:2019vvi}, 
and the 8~TeV CMS stop search with soft leptons, CMS-SUS-14-021~\cite{Khachatryan:2015pot}, 
with the corresponding EM results.

\section{Physics applications}

\subsection{Constraints on long-lived particles in the scotogenic model}

As first showcase for the usage of the new width-dependent results, we consider the scotogenic model~\cite{Ma:2006km,Kubo:2006yx}.  
Depending on the setup, this model features scalar or fermionic DM candidates. In either case, there can be long-lived charged scalars leading to HSCP, disappearing track, or displaced lepton signatures at the LHC. 

The scotogenic model supplements the SM by an additional SU(2) scalar doublet, $\Phi$, often referred to as the \emph{inert doublet}, and three%
\footnote{Variations of the model with less or more sterile neutrinos have been considered. To explain the observed oscillation in the active neutrino sector within the model, at least two sterile neutrinos need to be introduced. However, in the LHC phenomenology considered here maximally one of these states is involved, while the others are assumed to be heavy and not produced to significant amount.} 
sterile neutrinos, $N_n$. The new fields are taken to be odd under a new \Ztwo-parity, while the SM fields are even. 
The model Lagrangian is given by:
\begin{equation}
    \Lg=
        \Lg_\mathrm{SM}
        +\left| D^\mu \Phi \right|^2
        +\frac{i}{2}\bar N_n\slashed{\partial} N_n 
        -\left(\frac12 M_n \bar{N^c_n} N_n + i Y_{\alpha n} \bar L_\alpha \sigma_2 \Phi  N_n +\mathrm{h.c.}\right)
        -V(\Phi,H)
\end{equation}
where $\Lg_\mathrm{SM}$ is the Lagrangian of the SM, $M_n$ are the Majorana masses
of the right-handed neutrinos, and $Y$ is a $3\times 3$ complex matrix of Yukawa couplings. Finally, $V$ is the scalar potential
\begin{align}
    V(\Phi,H)= &
        \mu_1^2 |H|^2
        +\mu_2^2 |\Phi|^2
        +\lambda_1 |H|^4
        +\lambda_2 |\Phi|^4 
        +\lambda_3 |H|^2|\Phi|^2 \notag \\
        & +\lambda_4 |H^\dagger\Phi|^2
        +\frac{1}{2}\lambda_5 \left[(H^\dagger\Phi)^2+\mathrm{h.c.}\right]\,.
\end{align}
After electroweak symmetry breaking, where $\langle \Phi \rangle = 0$ in this model, the particle spectrum comprises five physical scalar states ($\hn$, $\Hn$, $\An$, $\Hp$) with masses:
\begin{align}
	m_{\hn}^2 &= \mu_1^2 + 3 \lambda_1 v^2\,,\notag \\
	m_{\Hn}^2 &= \mu_2^2 + \lambda_L v^2\,, \notag \\
	m_{\An}^2 &= \mu_2^2 + \lambda_S v^2\,,\notag \\
	m_{\Hp}^2 &= \mu_2^2 + \frac{1}{2} \lambda_3 v^2\,, 
\end{align}
where 
\begin{equation}
	v^2 = -\frac{\mu_1^2}{\lambda_1} \mbox{ and } \lambda_{L,S} = \frac{1}{2} \left( \lambda_3 + \lambda_4 \pm \lambda_5 \right)\,.
\end{equation} 
The first scalar, $h$, is even under the new \Ztwo-parity and identified with the observed SM-like Higgs boson, $m_h\simeq125$~GeV. The other scalars are \Ztwo-odd. Note that the scalar sector of the model has also been considered without the neutrino sector, often referred to as the inert doublet model (IDM). 

The presence of the new scalar and fermion fields provides a radiative generation of neutrino masses via the radiative seesaw mechanism~\cite{Ma:2006km,Kubo:2006yx}. Furthermore, the lightest \Ztwo-odd particle ($\Hn$ or $\An$, or the lightest of the sterile neutrinos, $N_1$) is stable and thus a natural DM candidate. In the following, we consider both options, \ie~scalar and fermionic DM candidates.

\subsubsection{Scalar (inert doublet) dark matter}
\label{sec:scalar}

The scalar sector of the scotogenic model provides two possible DM particles: the scalar $\Hn$ and the pseudoscalar $\An$; in the following we will identify the $\Hn$ as the DM candidate without loss of generality. The (collider) phenomenology of the scalar DM scenario is essentially the same as the one of the IDM\@.\footnote{Even if the sterile neutrinos are light, $m_{H^0}<m_{N_i}\lesssim m_{H^{\pm}},m_{A^0}$, their Yukawa couplings are too small to play a role for the collider phenomenology. However, the presence of sterile neutrinos can have an effect on the relic density, see the discussion in the main text.}

Regarding the relic density, 
for $m_{H^0}>m_W$, pair-annihilation of $\Hn$s into gauge bosons is so efficient that $\Omega h^2$ is typically much (up to 2 orders of magnitude) smaller than the observed value ($\Omega h^2=0.12$ \cite{Planck:2018vyg}).  One way to circumvent this conclusion  
is a small mass splitting between the inert scalars~\cite{LopezHonorez:2006gr,Goudelis:2013uca,Eiteneuer:2017hoh}. 
Indeed, for DM masses around 500--600 GeV one needs (sub-)GeV mass splittings (plus small $\lambda_L$) to achieve $\Omega h^2\sim 0.12$.  For such small mass differences, the $H^\pm$ can become long-lived and can be constrained by HSCP and disappearing track searches. 
The HSCP constraints were previously discussed in~\cite{Heisig:2018kfq} using \smoone.

In the scotogenic case, coannihilations with right-handed neutrinos close in mass to the $H^0$ can also help avoid DM under-abundance. As pointed out in \cite{Klasen:2013jpa,Avila:2021mwg},
these coannihilations tend to increase, rather than reduce, the freeze-out density and thus allow to satisfy the relic density constraint for DM masses well below 500 GeV. 
Finally, as studied in~\cite{Borah:2017dfn}, late decays of heavy right-handed neutrinos (happening after freeze-out of the scalar DM) can be an additional, non-thermal source of DM production and bring an initial under-abundance in agreement with the observed $\Omega h^2$. 

These considerations motivate us to consider long-lived charged scalars in the 100--900~GeV mass range and demonstrate how they are constrained by the width-dependent results in \smotwo. 
For the numerical analysis, we use the IDM implementation in \micromegas\ and carry out a random scan over $m_{H^0}$, $m_{H^{\pm}}-m_{H^0}$ and $m_{A^0}$, with the couplings fixed to $\lambda_2=0.01$ and $\lambda_L=10^{-10}$. 
The \micromegas-\smo\ interface~\cite{Barducci:2016pcb} (see also Appendix~\ref{app:MOinterface}) conveniently allows one to produce SLHA input files for \smo\ including masses, decay tables, and LHC cross sections computed with \calchep.  
The QNUMBER blocks are also automatically written by the interface. 

Before turning to the results, a few more comments on the parameter scan are in order. First, with the above choice of small $\lambda_2$ and $\lambda_L$, the production of inert scalars at the LHC is dominated by the SM gauge interaction arising from their kinetic terms; this gives conservative estimates of the LHC constraints. In principle values of $\lambda_2,\lambda_L\sim 10^{-2}$ are small enough to this end; 
the particularly tiny value of $\lambda_L=10^{-10}$ is chosen to have a vanishing Higgs portal coupling of the DM candidate, which is preferred in view of relic density constraints, see \eg~\cite{Borah:2017dfn}. 

Second, the overall mass spread within the inert scalars is constrained by the requirement that the quartic couplings be in the perturbative regime; here we simply require  $\lambda_{3,4,5} <4\pi$ as a rough bound. Moreover, and more importantly, the mass splittings are constrained by electroweak precision observables. For each point in the scan, we compute the oblique parameters $S$ and $T$ from eqs.~(24) and (25) in~\cite{Baak:2011ze} and demand that they fall within the 95\%~CL region of 
$S=0\pm 0.07$ and $T=0.05\pm 0.06$ with a correlation of $0.92$~\cite{ParticleDataGroup:2020ssz}. Very roughly, this limits $m_{\An}\lesssim 1.4\,m_{\Hn}$.

Third, regarding DM constraints, we require that $\Omega h^2<0.13$ from the standard thermal freeze-out calculation in \micromegas\ (assuming $\approx 10\%$ theory uncertainty from the tree-level calculation), and that the DM-nucleon scattering cross section rescaled by a factor $\Omega h^2/0.12$ evades the DM direct detection limits implemented in \micromegasversion. This, however, eliminates hardly any points. 

All in all, we sample 29k points that fulfill the above constraints in the region of small mass splittings, $m_{H^\pm}-m_{H^0}<0.5$~GeV, that gives long-lived charged scalars. 
The dominant decay modes of the $H^\pm$ are either into $\pi^\pm H^0$ 
(for $m_{H^\pm}-m_{H^0}>m_{\pi^\pm}=139$~MeV) 
or, for $m_{H^\pm}-m_{H^0}<m_{\pi^\pm}=139$~MeV, into $\ell^\pm H^0$ ($\ell=e,\mu$).%
\footnote{For mass differences $m_{H^\pm}-m_{H^0}$ below the QCD scale of around 1.5~GeV, the $H^\pm$ decay should be computed as decay into hadrons, $H^\pm \to \pi^\pm H^0$, instead of a 3-body decay into free quarks, $H^\pm \to u\bar d H^0$, via an off-shell $W$-boson~\cite{Chen:1995yu,Chen:1996ap,Chen:1999yf}. This is important for LLP studies, as $\Gamma(H^\pm \to \pi^\pm H^0) \gg \Gamma(H^\pm \to u\bar d H^0)$.}
Following \cite{Belyaev:2016lok,Belyaev:2020wok} (see also \cite{Belanger:2021lwd}), we add an effective $H^\pm H^0 \pi^\mp$ vertex 
\begin{equation}
  \frac{g^2 f_\pi}{4\sqrt{2} m_W^2} (p_{H^\pm} -p_{H^0}) \cdot p_{\pi^\mp}
\end{equation}
in the \calchep\ model file, with $f_\pi=0.13$~GeV the pion decay constant. This interaction arises from an effective, non pertubative $W$--$\pi$ Lagrangian ${\cal L}= (g f_\pi)/(2\sqrt{2}) W_\mu^+\partial^\mu \pi^-$ and gives a decay with of \cite{Belyaev:2020wok}
\begin{equation}
    \Gamma(H^\pm\to\pi^\pm H^0)=\frac{g^4 f_{\pi}^2}{64 \pi m_W^4}\Delta m^2\sqrt{\Delta m^2-m_{\pi^\pm}^2} \,,
\end{equation}
where $\Delta m = m_{H^{\pm}}-m_{H^0}$. 

\begin{figure}[t!]\centering
 \includegraphics[width=0.52\textwidth]{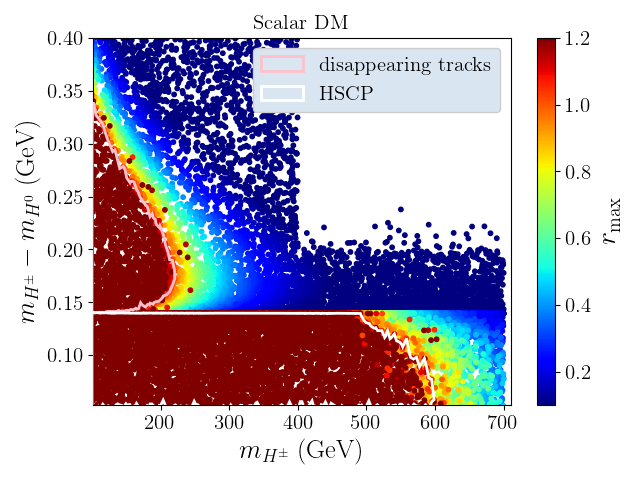}%
 \includegraphics[width=0.52\textwidth]{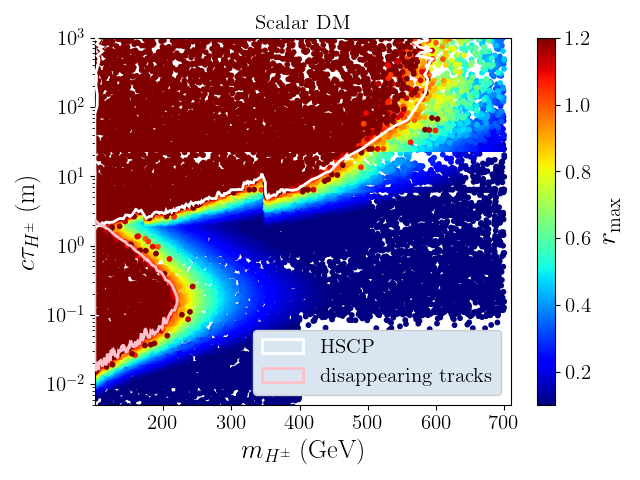}
 \caption{ \label{fig:ScotoScalarDM_rmax} 
\smo\ constraints on long-lived charged scalars in the scotogenic model with scalar DM, with the $H^\pm$ decaying either to $\pi^\pm H^0$ or to $\ell^\pm H^0$. The white line denotes the exclusion limit ($r_{\rm max}=1$) from HSCP searches, while the pink line denotes the exclusion limit from disappearing track searches; the value of $r_{\rm max}$ is shown in color. 
}
\end{figure}

The relevant long-lived signatures are disappearing tracks, which constrain lifetimes of ${\cal O}(1\,{\rm cm})$ to ${\cal O}(1\,{\rm m})$ for charged scalar masses of up to about 200~GeV, 
and HSCP signatures for lifetimes from about 2~m onward. 
The scan results are presented in Fig.~\ref{fig:ScotoScalarDM_rmax}, which shows the maximum $r$-value, $r_{\rm max}$, obtained from \smo, on the left in the plane of $m_{H^{\pm}}-m_{H^0}$ versus $m_{H^\pm}$, and on the right in the plane of mean decay length $c\tau_{H^\pm}$ versus $m_{H^\pm}$. 
The $r$-value is defined as the ratio of the theory prediction for a simplified model topology over the corresponding observed upper limit; points with $r\ge 1$ are therefore considered as excluded. 

The pink and white contours show the exclusion limits  ($r_{\rm{max}}=1$) from the disappearing track and HSCP searches, respectively. 
The HSCP limits are relevant for $c\tau_{H^\pm}\gtrsim 2$~m, corresponding to $m_{H^{\pm}}-m_{H^0}<m_{\pi}$. 
The sharp cut-off at the pion mass results from the rapid change in lifetime when the $H^\pm \to \pi^\pm H^0$ decay becomes kinematically allowed. The strongest HSCP constraints come from the 
13~TeV ATLAS analysis with $31.6$~fb$^{-1}$~\cite{Aaboud:2019trc}, the corresponding CMS analysis~\cite{CMS-PAS-EXO-16-036} in the \smo\ database having lower luminosity ($12.9$~fb$^{-1}$). For long lifetimes, $c\tau_{H^\pm}\gtrsim 100$~m, this ATLAS analysis excludes charged Higgs masses up to about 600 GeV. It is worth noting, however, 
that the 13~TeV analyses do not cover low masses $\lesssim160$~GeV as a result of the minimum cut on the reconstructed mass, see~Appendix~\ref{app:RecastHSCP}.  In this region, and for points close to the white contour up to $m_{H^\pm} \lesssim 500$~GeV, 
the exclusion comes from the 8~TeV CMS analysis~\cite{CMS:2013czn,CMS:2015lsu}.

For mean decay lengths ranging from few cm to about 2~m, corresponding to $m_{H^{\pm}}-m_{H^0}\simeq[0.14,\,0.35]$~GeV, the exclusion in Fig.~\ref{fig:ScotoScalarDM_rmax} comes from the ATLAS disappearing tracks search~\cite{Aaboud:2017mpt} with 36~fb$^{-1}$. This reaches up to $m_{H^\pm}\approx 220$~GeV for $c\tau_{H^\pm}\approx 20$~cm. 
We recall that here the EMs for scalar LLPs recasted by  \cite{Belyaev:2020wok,belyaev_2020_4288736} are used. 
In principle there is also the CMS disappearing track search \cite{CMS:2020atg} with 101 fb$^{-1}$ of data. For this analysis, however, simplified model results are available only for the chargino/neutralino hypothesis (fermionic LLPs). 
Since the disappearing track searches are implemented in the \smo\ database for specific spin assignments, see  Section~\ref{sec:llpresults}, by default the CMS  results are not applied to the scalar LLP scenario. This is, among other considerations, motivated by the fact that the disappearing track searches make use of hard jets originating from initial-state radiation. 

\begin{figure}[t!]\centering
 \includegraphics[width=0.6\textwidth]{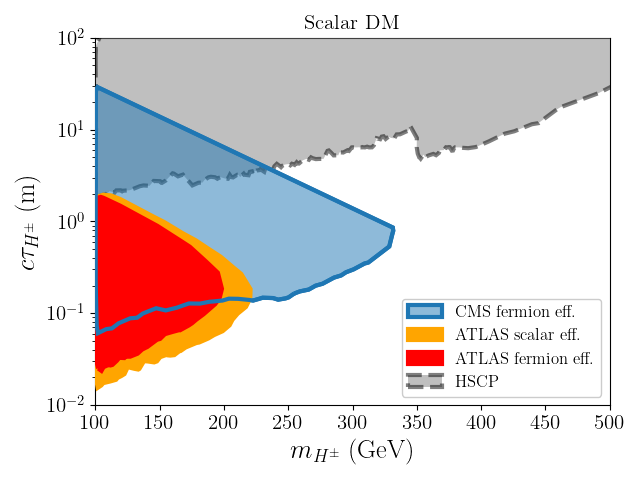}
 \caption{ \label{fig:DTcomparison} 
Comparison of exclusion limits for long-lived charged scalars as function of $H^\pm$ mass (in GeV) and lifetime (in m), using disappearing track EMs derived for either scalar or fermionic LLPs. 
}
\end{figure}

The picture that emerges when ignoring the spin dependence is shown in Fig.~\ref{fig:DTcomparison}. Indeed the CMS results  \cite{CMS:2020atg}, when applied to the scalar case, significantly extend the excluded region for decay lengths above about 10~cm (blue contour). For $c\tau_{H^\pm}\sim 1$~m, masses up to almost 350~GeV are excluded.  
Comparing the red and orange regions in Fig.~\ref{fig:DTcomparison}---the former being the ATLAS exclusion with EMs for the fermionic case, the latter the ATLAS exclusion with EMs for the scalar case---one sees that the effect of the spin dependence is rather small, roughly of the order of 10\% in the excluded mass. Moreover, the EMs for scalar LLPs actually exclude more than those for fermionic LLPs. It should thus be safe to apply the simplified model constraints from \cite{CMS:2020atg} to scalar LLPs, although one can expect some under-exclusion in this case. In any case, it would be desirable to have dedicated EMs for both spin choices.

Another comment is in order. The alert reader will have noticed the scattered red (excluded) points, outside the white and pink contours in Fig.~\ref{fig:ScotoScalarDM_rmax}. 
These points have $m_{A^0}-m_{H^\pm}<5$~GeV. For such small mass differences, the decay products of $A^0\to H^\pm$ and  $A^0\to H^0$ transitions are very soft and considered as invisible in \smo. Consequently, $A^0H^0\to H^\pm H^0+X_{\rm soft}$ and $A^0H^\pm\to H^\pm H^0+X_{\rm soft}$ production are treated as the same topology as $H^\pm H^0$ production (the $X_{\rm soft}$ being ignored), meaning their cross sections are added up in the simplified model decomposition. 
The same applies to $H^\pm H^\mp$ ($+X_{\rm soft}$) production. 
This is called \emph{mass compression} in \smo; the behaviour is controlled by the {\tt minmassgap} parameter, with {\tt minmassgap}$\,=5$~GeV being the default value. Since the cross sections are added up, the constraints typically become stronger once mass compression comes into play. There are, however, some differences between the disappearing tracks and 
HSCP results, which are worth explaining.

For the disappearing tracks search, results only exist for the direct production of one or two LLPs. Hence, charged scalars from the decay of the neutral inert states are not taken into account except for very small mass splittings where \smo\ applies the \emph{mass compression}. This explains why the exclusion reach increases for $m_{A^0}-m_{H^\pm}<5$~GeV. 

For the HSCP search~\cite{Aaboud:2019trc}, on the other hand, the database contains EM results for all topologies occurring in the IDM/scotogenic model. Hence, whether or not to apply the mass compression for small mass splittings should, in principle, not cause any difference. However, the EMs were produced using SUSY processes and it turns out that for two of the relevant 1-step topologies (concretely THSCPM8 and THSCPM11, see Fig.~\ref{fig:THSCPMdiag} and Table~\ref{tab:HSCPtopo} in Appendix~\ref{app:RecastHSCP}), the choices for the production modes and spins of the involved particles are different to the ones in the model considered here. These EMs tend to underestimate the efficiencies of the scalar case.\footnote{In the HSCP EMs, the spins and color charges are taken as arbirary.} For the direct HSCP production (topologies THSCPM1b and THSCPM2b), Drell-Yan production of a scalar was used, which matches the model considered here. As a consequence, when mass compression is applied for small mass differences, one gets a slightly stronger constraint from the EM results than without mass compression. 
Furthermore, the database also contains the UL results provided by the ATLAS collaboration for direct HSCP production. These ULs are slightly stronger than the limits obtained from the EMs for direct HSCP production and further strengthen the constraints in the case of mass compression with respect to the case where the individual topologies are combined.

\subsubsection{Fermionic (sterile neutrino) dark matter}

The lightest sterile neutrino is another phenomenologically viable DM candidate in the scotogenic model. For this choice, the observed relic density can be explained by the freeze-in mechanism~\cite{McDonald:2001vt,Asaka:2005cn,Hall:2009bx}. Considering freeze-in from the decays of inert scalars, to achieve $\Omega h^2\sim 0.12$ the respective Yukawa couplings should be of the order of~\cite{Molinaro:2014lfa,Hessler:2016kwm}
\begin{equation}\label{eq:y1}
\sqrt{|Y_{\alpha1}|^2} \sim  10^{-9}\left(\frac{10\,\mathrm{keV}}{m_{N_1}}\right)^{1/2}\left(\frac{\mu_2}{100\,\mathrm{GeV}}\right)^{1/2},
\end{equation}
where $\alpha=1,2,3$ runs over the three generations of leptons and $\mu_2$ represents the typical mass scale of the $Z_2$-odd scalars. To explain the observed oscillation of active neutrinos, the Yukawa couplings of the heavier sterile neutrinos are required to be considerably larger than the typical coupling for the lightest state, $10^{-5}<|Y_{\alpha2}|, |Y_{\alpha3}|<10^{-2}$~\cite{Hessler:2016kwm}.

Here, we are interested in the signatures expected from long-lived charged particles and focus on the scenario with  $m_{N_1}<m_{H^{\pm}}<m_{H^0},m_{A^0}<m_{N_2},m_{N_3}$ mass hierarchy. Given its small couplings, direct production of the sterile neutrinos can be neglected and, hence, ${N_2}$ and ${N_3}$ are phenomenologically irrelevant in our considerations.

In this scenario, $H^{\pm}$ can only decay into $N_1$ and a lepton. The respective decay width reads~\cite{Molinaro:2014lfa}
\begin{equation}\label{eq:decayHpm}
  \Gamma(H^{\pm}\rightarrow N_1 l_{\alpha}^{\pm})=\frac{m_{H^{\pm}}|Y_{\alpha 1}|^{2}}{16\pi}\left(1-\frac{m_{N_1}^{2}}{m^{2}_{H^{+}}}\right)^{2}\,.
\end{equation}
Requiring that the $N_1$ makes up for the observed DM abundance, eq.~\eqref{eq:y1}, the proper decay length of the $H^{\pm}$, $c\tau_{H^{\pm}}$, is of the order of
\begin{equation}\label{eq:ctauHpm}
  c\tau_{H^{\pm}}\sim 10\,\mathrm{m}\left(\frac{m_{N_1}}{10\,\mathrm{keV}}\right)\left(\frac{100\,\mathrm{GeV}}{m_{H^{\pm}}}\right)^2\,.
\end{equation}
Hence, in the relevant range of masses, the charged scalar is typically long-lived. When decaying outside the detector, ${H^{\pm}}$ leads to a HSCP signature while decays inside the tracker could be detected in searches for displaced leptons.

To showcase the sensitivity of \smotwo\ for this scenario, we perform a grid scan over $m_{H^{\pm}}$ and $m_{N_1}$, varying $m_{H^{\pm}}$ from 100 to 900\,GeV and $m_{N_1}$ from 1\,keV to 10\,GeV. The masses of the neutral \Ztwo-odd scalars are taken as $m_{H^0}=m_A^{0}=m_H^{\pm}+\Delta m$, with two choices for the mass splitting, $\Delta m=5$~GeV and $\Delta m=50$~GeV. Furthermore, we choose $\lambda_{L,S}=10^{-2}$. In this setup, the production of inert scalars at the LHC is dominated by SM gauge interaction arising from their kinetic terms. Finally, in our main results, we assume $Y_{\alpha 1}$ to have one non-zero entry only, such that $H^{\pm}$ decays to 100\% into one lepton flavor, either electrons or muons. While a non-trivial mixing, in general,  is subject to lepton flavor constrains, additionally, we consider the case of 50\% BR into muons and electrons for illustration. 

For a given particle spectrum, we compute the DM relic density arising from freeze-in production by solving the respective Boltzmann equation~\cite{Hall:2009bx}%
\footnote{We employ the commonly made approximations that the SM degrees of freedom do not vary during the freeze-in process and that scatterings and inverse decays could be neglected. We also neglect the contribution from late decays of the lightest scalar after its freeze out (often referred to as the superWIMP contribution). We checked that this contribution only becomes relevant outside the considered range of parameters, \ie~for larger DM and scalar masses.} taking into account the decays of all the inert scalars:
\begin{equation}\label{eq:fisol}
  \Omega h^2 \simeq 2.7\times 10^8\,m_{N_1} \int_0^{\infty}\!\mathrm{d} x\, \frac{K_1(x)}{x K_2(x)}\sum_{i}\frac{Y_{i}^\text{eq}(x)}{H(m_i/x)} \,\Gamma(i\to N_1\, l^\pm/\nu) \,,
\end{equation}
where $K_j$ denotes the modified Bessel functions of the second kind. The sum 
runs over the four inert degrees of freedom, $i \in\{H^\pm,H^0,A^0\}$, $Y_{i}^\text{eq}(x)$ denotes their comoving number density in thermal equilibrium, and $\Gamma (i\to N_1\, l^\pm/\nu)$ the partial width for their decays into $N_1$ and a charged lepton or neutrino. The latter is taken from~\cite{Molinaro:2014lfa}. The Hubble parameter, $H$, is evaluated at the temperature $T=m_i/x$. 

Contrary to the scalar DM case, here we have less freedom for generating the DM; we therefore assume that the $N_1$ saturates the relic density constraint through freeze-in.
As the partial decay widths involved in eq.~\eqref{eq:fisol} are all proportional to $|Y_{\alpha 1}|^{2}$, we solve for $\Omega h^2=0.12$ for this Yukawa coupling and compute the corresponding lifetime of the $H^\pm$ for a given set of masses. Note that for the mass splittings considered here, $H^0$ and $A^0$ promptly decay into $H^\pm$ such that  their direct decays into $N_1$ can safely be neglected.

\begin{figure}[t!]\centering
 \includegraphics[width=0.52\textwidth]{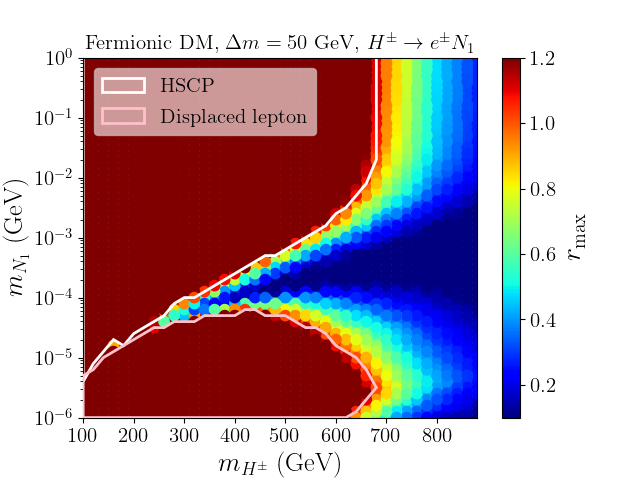}%
 \includegraphics[width=0.52\textwidth]{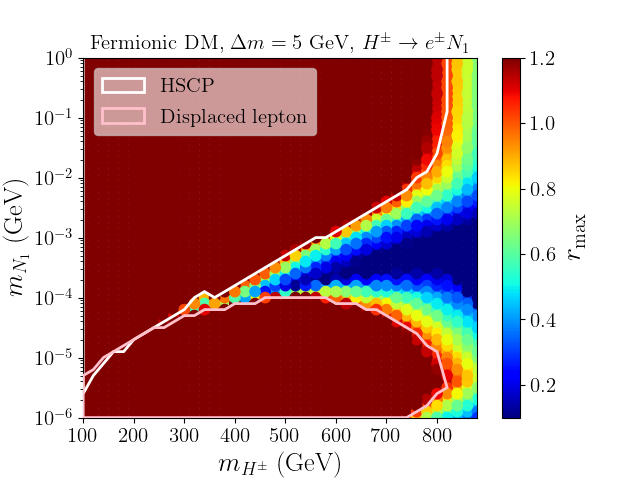}\\
 \includegraphics[width=0.52\textwidth]{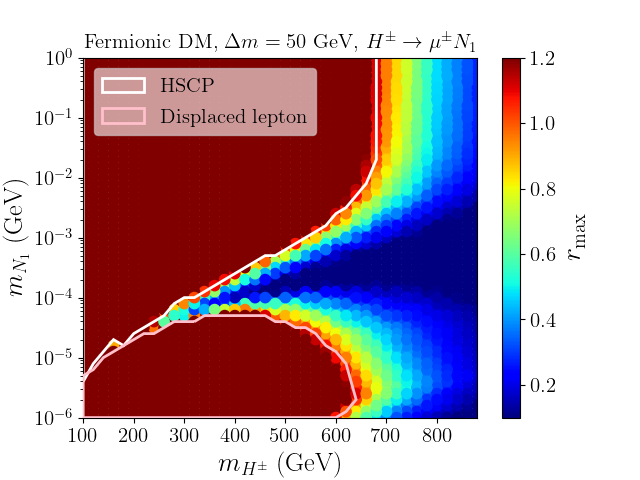}%
 \includegraphics[width=0.52\textwidth]{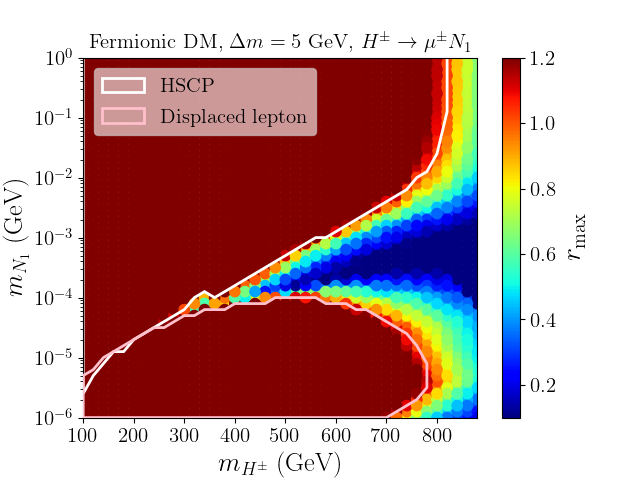}
 \caption{ \label{fig:ScotoFermionDM_Rmax} 
\smo\ constraints on long-lived charged scalars in the scotogenic model with fermionic DM, where $H^\pm\to \ell^\pm N_1$. On the left $\Delta m \equiv m_{A^0,H^0}-m_{H^\pm}=50$~GeV, on the right  $\Delta m = 5$~GeV. The white lines denote the exclusion limit ($r_{\rm max}=1$) from HSCP searches, while the pink lines denote to exclusion limit from displaced lepton searches; the value of $r_{\rm max}$ is shown in color.
The upper panels are for $\ell^\pm=e^\pm$, the lower panels for $\ell^\pm=\mu^\pm$ with 100\% branching ratio. 
}
\end{figure}

As in the previous subsection, we compute the production cross sections of the \Ztwo-odd scalars with \calchep\ utilizing the \micromegas-\smo\ interface, and then evaluate the LHC constraints with \smotwo.
Figure~\ref{fig:ScotoFermionDM_Rmax} shows the obtained $r_{\rm max}$ values in the $m_{H^{\pm}}$ vs.\ $m_{N_1}$ plane as well as the \smo\  exclusion lines (contours of $r_{\rm max}=1$) resulting from the HSCP and displaced lepton searches, for the two choices of $\Delta m=m_{A^0,H^0}-m_{H^\pm}$ and the $H^\pm$ decaying either 100\% into $e^\pm N_1$ or 100\% into $\mu^\pm N_1$. 

The 13~TeV ATLAS search for displaced leptons~\cite{ATLAS:2020wjh} provides sensitivity in the region $c\tau\lesssim 1$~m. The HSCP limits are strongest in the detector-stable limit, while for decay lengths of  ${\cal O}$(1~m) they suffer from the exponentially suppressed fraction of particles traversing the detector. The two exclusion regions have a minor overlap for small $m_{H^\pm}$, \ie\ a large production cross section. 
The relevant HSCP analyses are the 8~TeV CMS searches~\cite{CMS:2013czn,CMS:2015lsu}, and the 13~TeV ATLAS search~\cite{Aaboud:2019trc}. As in the previous section, the 8~TeV analysis constrains the region up to $m_{H^{\pm}}\lesssim 160$~GeV and in the case $\Delta m=5$~GeV,  $c\tau_{H^{\pm}}\sim\mathcal{O}(5~\mathrm{m})$ up to 400~GeV.

In the case $\Delta m=50$~GeV (left panels in Fig.~\ref{fig:ScotoFermionDM_Rmax}), both the HSCP and displaced lepton searches exclude $H^{\pm}$ masses up to about 700~GeV. The displaced lepton limits are slightly stronger for decays into electrons than for decays into muons. For smaller mass differences, $\Delta m=5$~GeV (left panels), the limits are stronger, reaching up to $m_{H^{\pm}}\approx 800$~GeV. This is due to the fact that 
the $H^\pm$ can be produced in $pp\to H^+H^-$, $H^\pm A^0$ and $H^\pm H^0$ channels, as well as from $pp\to A^0H^0$ with the neutral scalars decaying to the charged one, and the total $H^\pm$ signal thus increases with decreasing $\Delta m$. For the HSCP constraints, all these different signal contributions can be summed up thanks to the large variety of home-grown EMs (see however the caveat in case of mass compression explained in Section~\ref{sec:scalar}). 
In the case of displaced leptons, similar to the case of disappearing tracks considered in Section~\ref{sec:scalar}, simplified model results  only exist for the direct production of the LLP; $H^\pm$ originating from $H^0$ or $A^0$ decays are therefore ignored unless \smo's mass compression sets in.

Let us now turn to the question of lepton flavor. So far, 
we assumed 
that $H^{\pm}$ decays to 100\% into one lepton flavor, concretely either electrons or muons.\footnote{Results for displaced taus are also included in the \smotwo\ database.} As $m_{H^{\pm}}\gg m_e,m_\mu$, both scenarios provide the same $H^{\pm}$ decay widths. The HSCP constraints, depending only on mass and lifetime of the LLP, stay agnostic to the type of decay products. The constraints from the displaced lepton search are, however,  sensitive to the lepton flavor. Indeed, the slight difference in the displaced lepton limits in the upper and lower panels of Fig.~\ref{fig:ScotoFermionDM_Rmax} arises due to the larger acceptance $\times$ efficiency values in signal region SRee (two displaced electrons) as compared to SRmm (two displaced muons). 

\begin{figure}[t!]\centering
  \includegraphics[width=0.52\textwidth]{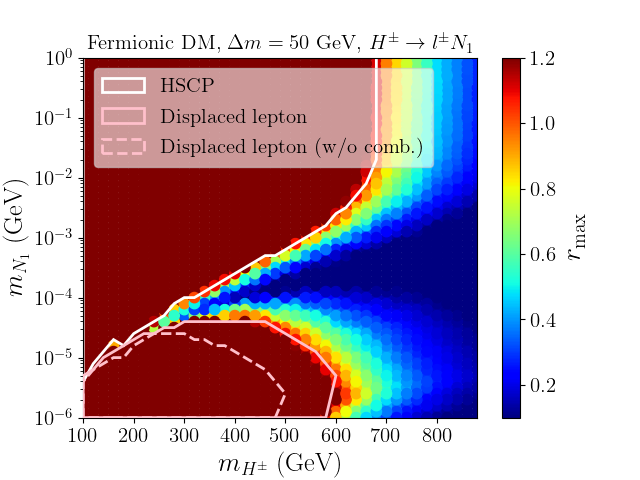}%
  \includegraphics[width=0.52\textwidth]{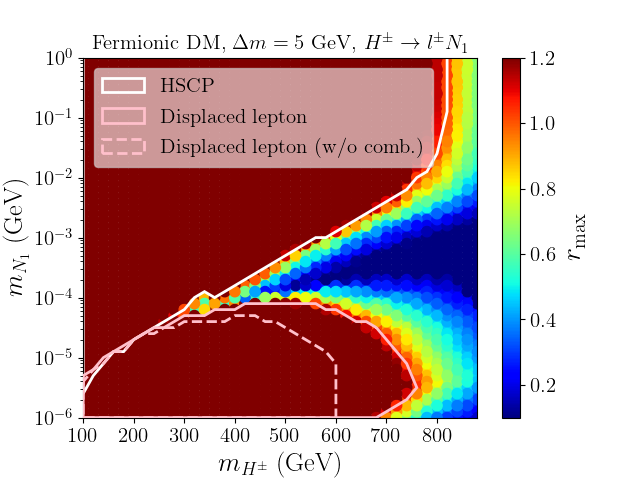}
 \caption{ \label{fig:ScotoFermionDM_5050lep} 
As Fig.~\ref{fig:ScotoFermionDM_Rmax} but for $H^\pm\to \ell^\pm N_1$ decays with equal probabilities for $\ell=e,\,\mu$ (50\% BRs). Here, \smo\ is run with {\tt combineSRs=True} in order to combine the contributions in the SRee and SRmm signal regions of the ATLAS displaced lepton search~\cite{ATLAS:2020wjh}; this is made possible by the  statistical model provided by the collaboration. For comparison, the pink dashed lines show the results without SR combination. }
\end{figure}

Any non-trivial structure in $Y_{\alpha 1}$, while subject to lepton flavor violation constraints, will lead to a mix of lepton flavors in the $H^\pm \to \ell^\pm N_1$ decays. Staying with $\ell=e,\mu$ for simplicity, the worst case is BR($H^\pm \to e^\pm N_1$)=BR($H^\pm \to \mu^\pm N_1$)=50\%, for which only 25\% of the total $H^+H^-$ production 
gives displaced leptons of the same flavor. This would considerably reduce the constraining power. 
Fortunately, the ATLAS collaboration published the statistical model for the analysis in JSON format, which allows for the combination of the SRee and SRmm signal regions in this case.\footnote{With the EM results provided by ATLAS, contributions in the mixed-flavor signal region SRem can only be included in addition in case of displaced taus.}  
To demonstrate the usefulness of combining the SRs in the displaced lepton search, we consider in Fig.~\ref{fig:ScotoFermionDM_5050lep} the case of maximal mixture between the first two generations, such that 
$H^\pm\to \ell^\pm N_1$ decays give $\ell=e,\,\mu$ with equal probabilities. The full pink contour shows the exclusion reach when 
SR combination is turned on ({\tt combineSRs=True}); this has to be compared to the dashed pink contour which represents the exclusion line  without SR combination.

Before concluding this subsection, we note that the case of small DM masses is also constrained from cosmological observations independent of the scalar sector. In particular, 
a recent reinterpretation of warm DM constraints from the Lyman-$\alpha$ forest~\cite{Palanque-Delabrouille:2019iyz} in the freeze-in scenario excludes masses below 15 (3.5) keV~\cite{Decant:2021mhj} under nominal (conservative~\cite{Garzilli:2019qki}) assumptions.

\subsection{Constraints on electroweak-inos in the MSSM}

Our second showcase is the EW-ino sector of the MSSM\@. As the lifetime of the lighter chargino $\tilde \chi_1^\pm$ can span a wide range of values, from prompt decays to decay lengths of several cm, this serves to illustrate the simultaneous usage of prompt and LLP results in \smotwo. 

To cover the parameter space of the EW-ino sector, 
we perform a random scan over the relevant Lagrangian parameters, \ie\ the bino and wino mass parameters $M_1$ and  $M_2$, the higgsino mass parameter $\mu$, and $\tan\beta=v_2/v_1$. Concretely, we vary 
\begin{eqnarray}
    10 \mbox{ GeV} < & M_1 & < 3 \mbox{ TeV}, \nonumber\\
    100 \mbox{ GeV} < & M_2 & < 3 \mbox{ TeV}, \nonumber\\
    100 \mbox{ GeV} < & \mu & < 3 \mbox{ TeV}, \\
    5 < & \tan \beta & < 50. \nonumber
\end{eqnarray}
The other SUSY breaking parameters are fixed to 10~TeV.\footnote{We assume that stop parameters can always be adjusted such that $m_h\approx 125$~GeV without influencing the EW-ino sector.} 
The lower limits on $M_2$ and $\mu$ are chosen so to avoid the LEP constraints on light charginos.
The scan 
consists of close to 100k points, generated randomly within the parameter intervals above. The mass spectra and decay tables are computed with \textsc{SoftSusy}~4.1.12~\cite{Allanach:2001kg,Allanach:2017hcf}, which includes the  $\tilde\chi^\pm_1\to \pi^\pm\tilde\chi^0_1$ decay calculation following~\cite{Chen:1995yu,Chen:1996ap,Chen:1999yf}
(see also \cite{Goodsell:2020lpx}) for small mass differences below about 1.5~GeV. Cross sections 
are computed first at leading order (LO) with \textsc{Pythia\,8}~\cite{Sjostrand:2007gs,Sjostrand:2014zea}, and reevaluated at next-to-LO with \textsc{Prospino}~\cite{Beenakker:1996ed} for all points which have $r_{\rm max}>0.7$ with the LO cross sections. 
Furthermore, we take the lightest neutralino $\tilde \chi_1^0$ to be the lightest supersymmetric particle (LSP) and a DM candidate;  
its relic density and scattering cross sections on nucleii are computed with  \micromegas~v5.2.7a~\cite{Belanger:2013oya,Belanger:2018ccd}. 

\begin{figure}[t!]\centering
 \includegraphics[width=0.8\textwidth]{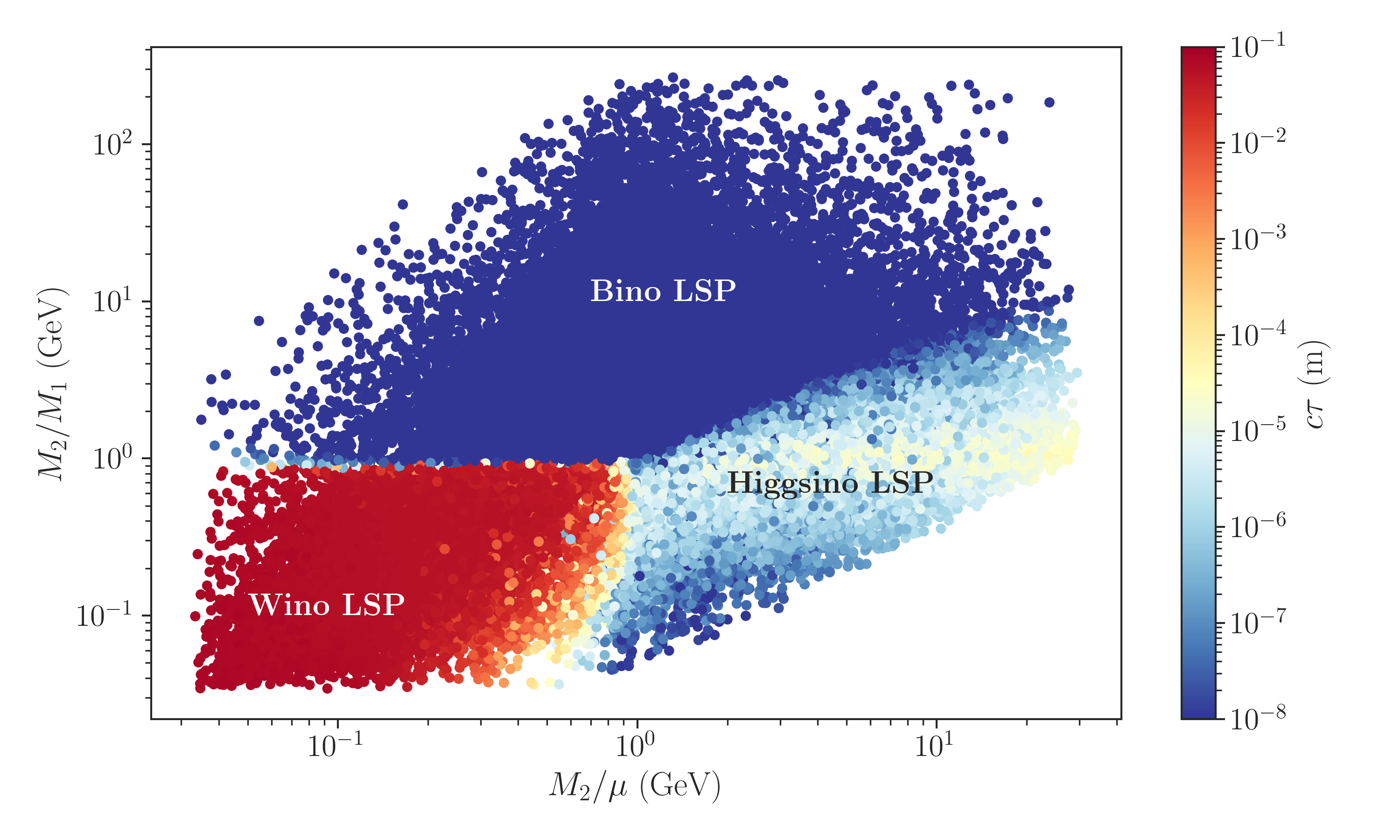}
 \caption{Mean decay length $c\tau$ of the lighter chargino,  $\tilde{\chi}_1^\pm$, for the scan described in the text. The red points at the lower left corner correspond to the wino scenario ($M_2 \ll \mu,\,M_1$) and have the largest decay lengths (up to 8 cm). The light blue to yellowish points on the right side correspond to the higgsino scenario ($\mu \ll M_2,M_1$), resulting in decay lengths of $\sim 0.1$~mm. The dark blue points represent prompt decays. \label{fig:charginoLifetime} 
}
\end{figure}

The mean decay length of the $\tilde \chi_1^\pm$ is shown in Fig.~\ref{fig:charginoLifetime}. As can be seen, a large fraction of the scan points feature prompt decays and can thus be tested by the conventional $\etmiss$ SUSY searches. On the one hand, the wino LSP case ($M_2 \ll \mu,\,M_1$) can result in lifetimes as large as a few centimeters, which can be tested by the ATLAS and CMS disappearing track searches. The higgsino LSP scenario ($\mu \ll M_2,M_1$) can also lead to non-prompt decays, \cf~the light blue and yellowish points in Fig.~\ref{fig:charginoLifetime}, but these remain in the sub-mm regime and are thus not probed by the current LLP results. 

\begin{figure}[t!]\centering
 \includegraphics[width=0.8\textwidth]{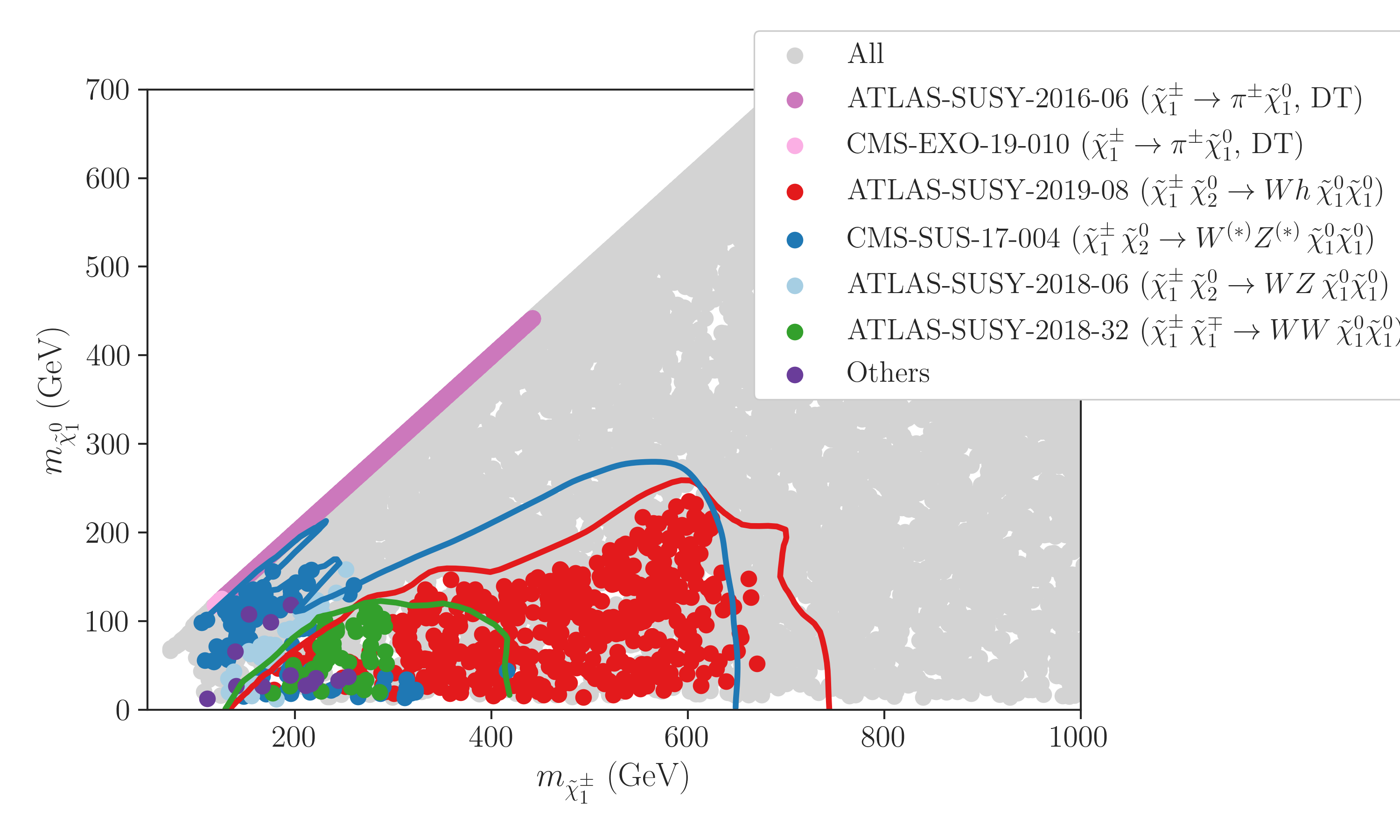}
 \caption{ \label{fig:EWexclusionMassMass} 
Constraints on EW-inos in the MSSM, from the scan explained in the text. The points in colour are excluded by \smo, with the colour denoting the analysis that gives the highest $r$-value (see legend). The simplified model exclusion lines from the respective ATLAS and CMS publications are also shown for comparison.}
\end{figure}

Figure~\ref{fig:EWexclusionMassMass} shows the points excluded by the LHC searches in the \smotwo\ database in the $m_{\tilde \chi_1^\pm}$ versus $m_{\tilde \chi_1^0}$  plane. 
The color of each excluded point denotes the most constraining analysis, that is the analysis giving the highest $r$-value, $r_{\rm max}$. 
As we can see, points in the mass-degenerate region are excluded by the ATLAS~\cite{Aaboud:2017mpt} and CMS~\cite{CMS:2020atg} disappearing track searches (light and dark pink points). These points correspond to the wino LSP  scenario. The compressed region, where $m_{\tilde \chi_1^\pm} - m_{\tilde \chi_1^0} \lesssim m_W$, is mostly tested by the CMS searches~\cite{Sirunyan:2018ubx,Aad:2019vvi} in final states with off-shell $W$ and $Z$ bosons (dark blue points). For larger mass differences, the decay $\tilde \chi_2^0 \to \tilde \chi_1^0 h$ starts to become kinematically accessible and the branching ratio BR$(\tilde \chi_2^0 \to \tilde \chi_1^0 h)$ starts to increase while BR$(\tilde \chi_2^0 \to \tilde \chi_1^0 Z)$ is reduced.
In this transition region, the effective cross section for the $\tilde \chi_1^\pm \tilde \chi_2^0 \to W Z + \etmiss$ simplified model is reduced and constraints from chargino-pair production~\cite{Aad:2019vnb} become more relevant (green points). For larger chargino masses, the decay $\tilde \chi_2^0 \to \tilde \chi_1^0 h$ becomes dominant and constraints from the $Wh+\etmiss$ final state \cite{Aad:2019vvf} kick in (red points). Finally, other ATLAS and CMS searches in the $WZ+\etmiss$ final state can be more constraining for a few points in parameter space; for simplicity, they are grouped as ``Others'' in Fig.~\ref{fig:EWexclusionMassMass}.

\begin{figure}[t!]\centering
 \includegraphics[width=0.8\textwidth]{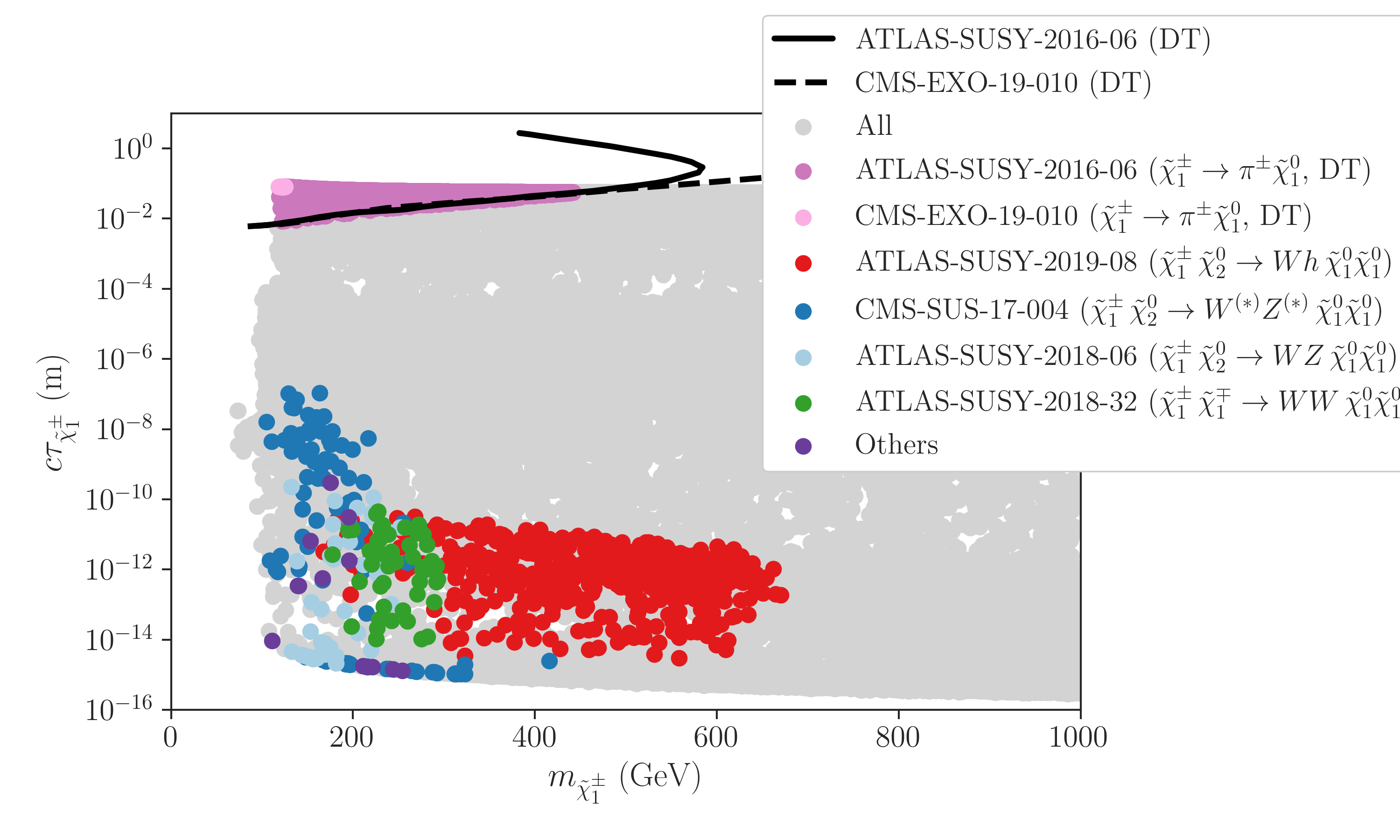}
 \caption{ \label{fig:EWexclusionMassCtau} 
As Fig.~\ref{fig:EWexclusionMassMass} but in the plane of $\tilde\chi^\pm_1$ mass versus mean decay length.  The simplified model exclusion lines from the ATLAS and CMS disappearing tracks searches are also shown for comparison.}
\end{figure}

It is also instructive to consider the same points plotted in the plane of chargino mass vs.\ mean decay length, shown in Fig.~\ref{fig:EWexclusionMassCtau}. As discussed above, in the wino LSP case ($M_2 \ll M_1,\mu$), the lighter chargino is long-lived and can be constrained by the disappearing track searches, which are sensitive down to decay lengths of $\sim 1$~cm.

\begin{figure}[t!]\centering
 \includegraphics[width=0.8\textwidth]{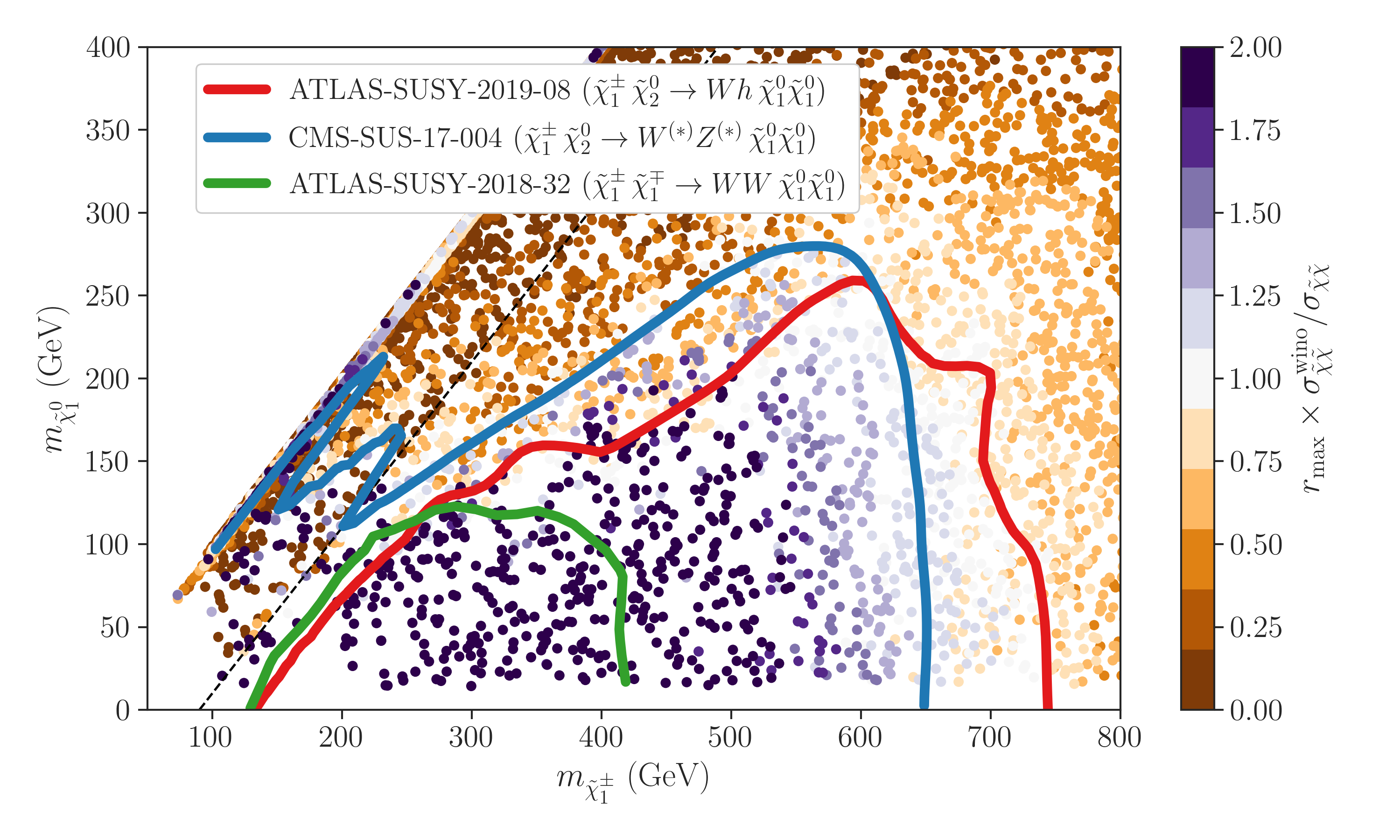}
 \caption{ \label{fig:EWallowedXsec} 
\smo-allowed scan points in the plane of $m_{\tilde\chi^0_1}$ versus $m_{\tilde\chi^\pm_1}$; the colour code shows the highest $r$-value rescaled by the reference cross sections for pure wino production.}
\end{figure}

In Figs.~\ref{fig:EWexclusionMassMass} and \ref{fig:EWexclusionMassCtau}, we also display the exclusion curves published by ATLAS and CMS for the corresponding simplified models. For the case of disappearing track searches, the exclusion obtained with \smo\ agrees very well with the `official' 
exclusion curves from the collaborations. 
For the prompt searches, however, \smo\ seems to underestimate the reach as compared to the exclusion limits from the ATLAS and CMS collaborations.
Generally, the exclusion obtained with \smo\ tends to be conservative, since it is limited by the simplified models included in the database. The main reason behind the supposed under-exclusion in Fig.~\ref{fig:EWexclusionMassMass} is, however, the fact that the ATLAS and CMS mass limits assume a pure bino $\tilde\chi^0_1$ and pure wino $\tilde \chi_1^\pm$ and $\tilde \chi_2^0$, the latter assumption maximising the production cross section. Within the MSSM, this is only approximately valid if $M_1 \ll M_2 \ll \mu$. For general parameters as in our scan, the production cross sections will typically be smaller than the ones assumed by the collaborations, thus reducing the excluded region. 
To illustrate this point, we plot in Fig.~\ref{fig:EWallowedXsec}
all the non-excluded points in the $m_{\tilde \chi_1^\pm}$ vs.\ $m_{\tilde \chi_1^0}$ plane. The color coding shows the $r_{\rm max}$ value obtained for each (allowed) point rescaled by the pure wino production cross section ($\sigma_{\tilde \chi \tilde \chi}^{\rm wino}$). In other words, it shows which would be the $r_{\rm max}$ value if the production cross sections for $\tilde \chi \tilde \chi$, with $\tilde \chi = \tilde \chi_1^\pm,\tilde \chi_2^0$, were the ones assumed in the ATLAS and CMS limit plots. As we can see, almost all non-excluded points that fall within the ATLAS and CMS exclusion curves become excluded ($r_{\rm{max}} \times \sigma_{\tilde \chi \tilde \chi}^{\rm wino}/\sigma_{\tilde \chi \tilde \chi} \ge 1$) once their cross sections are rescaled in this way.

\subsubsection*{LHC versus dark matter constraints}

Let us next consider possible DM constraints for the EW-ino scenario, and their complementarity with the LHC constraints. To this end, we assume a standard cosmological history, so the $\tilde\chi^0_1$ relic abundance, $\Omega h^2$, is given by the usual WIMP freeze-out calculation. Furthermore, we allow for the $\tilde\chi^0_1$ to make up for just a fraction of the observed DM abundance, which may include contributions from other, non-MSSM particles. Therefore we impose only an upper bound of $\Omega h^2 < 0.13$ (assuming again $\approx 10\%$ theory uncertainty from the tree-level calculation) and rescale the DM-nucleon cross section that enters the DM direct detection constraints by a factor $\Omega h^2/0.12$.

\begin{figure}[t!]\centering
 \includegraphics[width=0.9\textwidth]{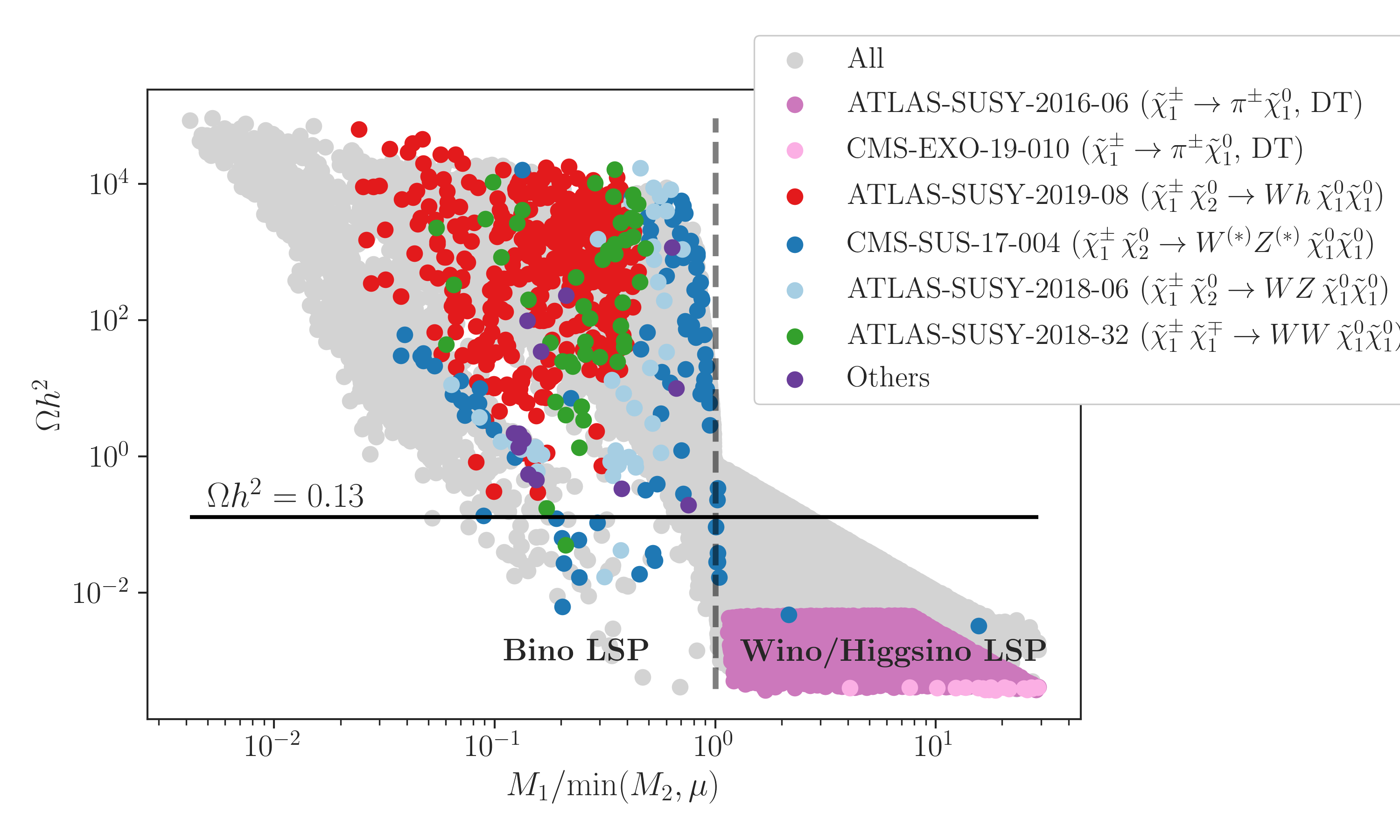}
 \caption{ Comparison of LHC and the DM relic abundance constraint for the EW-ino scenario; see text for details. The regions with a pure bino DM corresponds to $M_1/{\rm min}(M_2,\mu)\ll 1$ and with a wino/higgsino DM to $M_1/{\rm min}(M_2,\mu)\gg 1$, as indicated by the dashed line. The upper bound on the LSP relic abundance is shown by the solid line.\label{fig:EWexclusionDM} 
}
\end{figure}

It is well known that in the EW-ino scenario, the bino LSP case ($M_1 \ll M_2,\,\mu$) leads to a DM overabundance and would be excluded by the considerations above, while the wino and higgsino LSP cases lead to an under-abundance for masses below $\sim1$~TeV. From the discussion above we know that the LHC constraints from prompt searches are more stringent for the bino LSP case, with wino-like $\tilde \chi_1^\pm$ and $\tilde \chi_2^0$. Wino LSP cases, on the other hand, are constrained by disappearing track searches.
To compare the LHC and DM constraints, we show in Fig.~\ref{fig:EWexclusionDM} the $\tilde\chi^0_1$ relic abundance as a function of the ratio $M_1/{\rm min}(M_2,\mu)$. 
Values of this ratio much smaller than 1 correspond to the bino LSP scenario, while values much larger than one correspond to the wino/higgsino LSP case. As we can see, the points excluded by prompt searches are almost entirely restricted to the bino LSP region, which is already excluded by the relic abundance constraint. The exception are a few points with a mixed LSP leading to a small relic.  
On the other hand, the disappearing track searches are sensitive to the wino LSP case with $\Omega h^2\lesssim 10^{-2}$.

\begin{figure}[t!]\centering
 \includegraphics[width=0.9\textwidth]{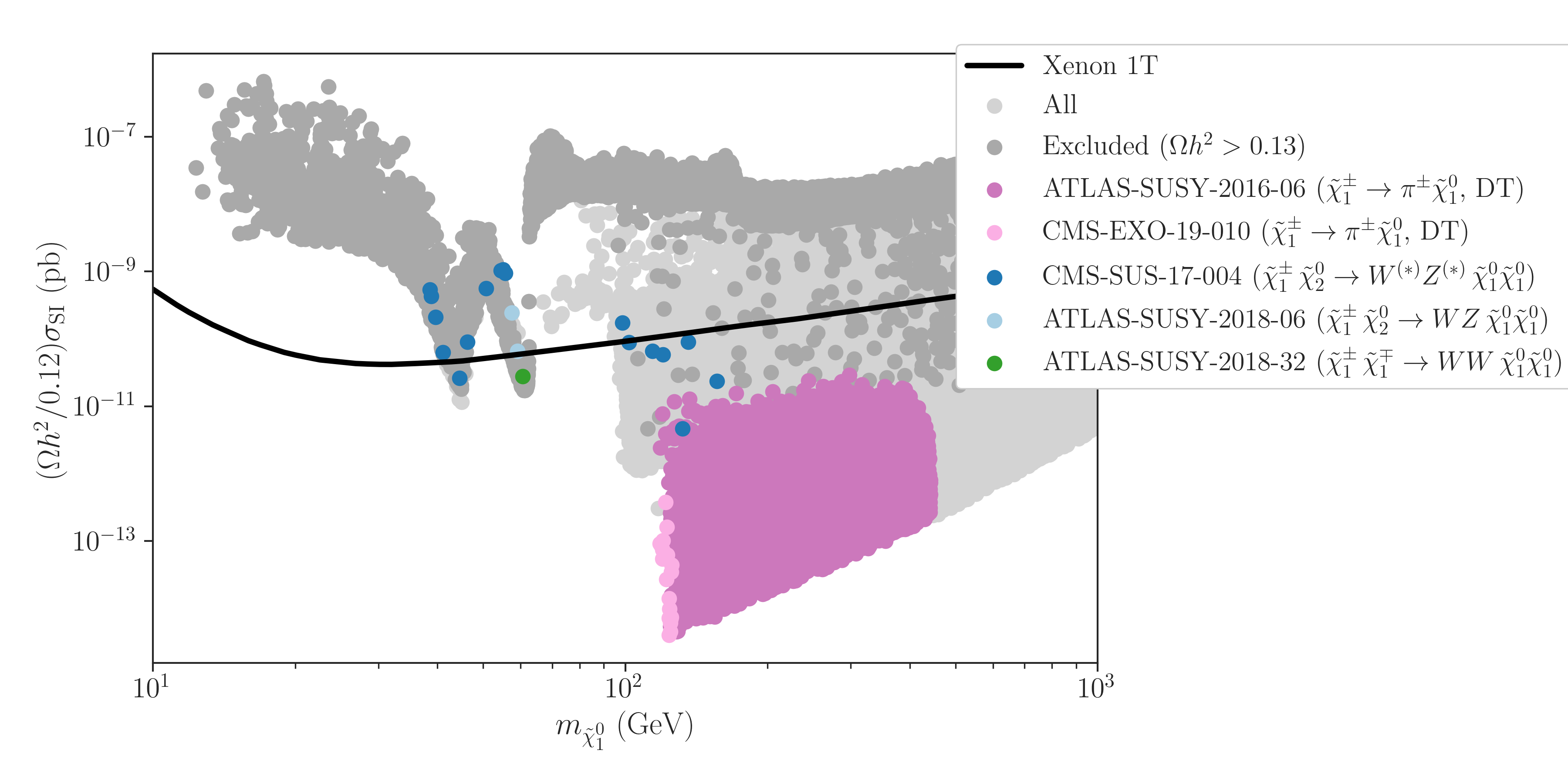}
 \caption{ \label{fig:EWexclusionDD} 
Comparison of relic density, DM direct detection and LHC constraints for the EW-ino scenario; see text for details. 
}
\end{figure}

Finally, in Fig.~\ref{fig:EWexclusionDD} we show the effective (\ie~rescaled) $\tilde\chi^0_1$-nucleon scattering cross section as a function of the $\tilde\chi^0_1$ mass. Points with too high a relic abundance ($\Omega h^2 > 0.13$) are shown in dark grey, while the points with $\Omega h^2 < 0.13$ but excluded by LHC results are shown in color (light and dark shades of blue, green and pink). The 90\% CL direct detection limit from the Xenon1T experiment \cite{XENON:2018voc} is shown as black line. 
Once again, just a handful of points excluded by the prompt searches can evade the DM (relic plus direct detection) constraints.
The pure wino case, however, leads to a suppressed relic abundance 
and, consequently, to a small effective direct detection cross section, thus evading the DM constraints. These points are only constrained by the disappearing track searches, as shown by the pink points in Fig.~\ref{fig:EWexclusionDD}.

\section{Conclusions}

Version~2 of \smo\ features a more detailed and more flexible  description of simplified model topologies. 
Concretely, through the introduction of a \emph{particle class}, the simple list of BSM particle masses used in \smoone\ has been replaced by full \emph{objects}, which can carry attributes such as mass, width, spin, electric charge, etc. 
This enables a refined treatment of, \eg, spin-dependent results. Moreover, and more importantly for this paper, it enables the inclusion of a large variety of searches for LLPs in the form of width-dependent results. 
Given an input model, \smotwo\ can thus simultaneously provide prompt and long-lived results in the same run. 
The LLP searches currently implemented include searches for HSCPs, disappearing tracks, displaced/non-prompt jets and displaced leptons. In total, results from 62 ATLAS and CMS searches at 13 TeV are implemented in the current database, 7 of which are for LLPs. 

We demonstrated the physics capabilities of \smotwo\ analysing the constraints on long-lived charged scalars in the scotogenic model with either scalar or fermionic DM. 
In the former case, LLPs arise for sub-GeV mass splitting with the scalar DM candidate. Considering the charged scalar as the next-to-lightest \Ztwo-odd state, we found that HSCP searches constrain its mass up to around 500--600~GeV for mass splittings below the pion threshold. This covers the range  favoured by DM considerations for such small mass splittings. 
For mass splittings slightly above $m_{\pi^\pm}$, resulting in decay lengths of the order of 0.1--1~m, we found that disappearing track searches from ATLAS with 36~fb$^{-1}$  (CMS with 101~fb$^{-1}$) 
can exclude charged scalar masses up to around 220 (350)~GeV, depending on the lifetime. 

For fermionic DM in the scotogenic model, the measured relic density can be explained by freeze-in production. The smallness of the required Yukawa coupling of the lightest sterile neutrino naturally renders the next-to-lightest \Ztwo-odd state long-lived in a large region of the cosmologically valid parameter space where the DM mass ranges from the keV to the GeV scale. Besides  the searches for HSCP that provide sensitivity to large lifetimes (corresponding to DM masses in the MeV to GeV range) we applied constraints from searches for displaced leptons that are most sensitive towards smaller DM masses, below a few tens of keV. 
Both types of searches can exclude long-lived charged scalars up to about 800~GeV in mass in this scenario. For the displaced lepton search, we also demonstrated the importance of signal region combination, as enabled by the statistical model provided by ATLAS for this analysis. 

As a second showcase we analysed the constraints on the EW-ino sector of the MSSM, detailing the interplay of prompt and long-lived searches, as well as the interplay of collider and DM constraints.  
We demonstrated to what extend the ATLAS and CMS EW-ino mass limits, which assume a pure bino $\tilde\chi^0_1$ and pure wino $\tilde \chi_1^\pm$ and $\tilde \chi_2^0$ (the latter in order to  maximise the production cross section) are weakened in the general case. One aspect is that  $\tilde\chi^0_2$ decays into $\tilde\chi^0_1Z$ or $\tilde\chi^0_1h$ compete with each other if both are kinematically allowed. More importantly, however, for general EW-ino mass parameters as in our scan, the production cross sections are typically  smaller than the ones assumed by the collaborations, thus reducing the excluded region. 

Regarding the interplay with DM constraints, we showed that EW-ino scenarios excluded by prompt searches are almost entirely restricted to the bino LSP region and characterised by a significant DM overabundance. Only few points with a mixed LSP, leading to a small relic, escape this conclusion and evade also the DM direct detection bounds.  
The disappearing track searches, on the other hand, are sensitive to the wino LSP case with $\Omega h^2\lesssim 10^{-2}$. In this case, the suppressed relic abundance also leads a small effective direct detection cross section,  evading the Xenon1T limit by order(s) of magnitude, to the effect that this scenario is best tested at colliders.

\smotwo\ is publicly available on GitHub~\cite{smodels:wiki} and can serve the whole community for fast testing of LHC constraints for BSM models that feature a \Ztwo-like symmetry. As \Ztwo-like symmetries are prevalent in DM models, \smo\ is also interfaced from \mbox{\micromegas}. A lot more work is foreseen to further extend and improve the usage of simplified model results and both, the \smo\ code and the database, will continue to evolve.

\acknowledgments

We thank the ATLAS and CMS collaborations for the provision of a plethora of simplified model results in readily usable format on HEPData. We are particularly grateful for the large number of  efficiency map results supplementing the upper limit maps, and for the full likelihoods in JSON format.
Furthermore, we wish to thank Laura Jeanty for helpful discussions on the ATLAS HSCP analysis, and A.~Belyaev et al.\ for providing dedicated efficiency maps from \cite{Belyaev:2020wok} for our purpose on Zenodo.

The work of G.A.\ and S.K.\ was supported in part by the IN2P3 master project ``Th\'eorie -- BSMGA''\@. 
J.H.~acknowledges support from the DFG via the Collaborative Research Center TRR 257 and the F.R.S.-FNRS (Charg\'e de recherches). 
Su.K.~is supported by the Austrian Science Fund through  Elise-Richter grant project number V592-N27\@. 
The research of H.R.G.\ is funded by the Italian PRIN grant 20172LNEEZ\@.
A.L.~acknowledges support from FAPESP (project grant 2018/25225-9).
W.W.~acknowledges financial support from the Universit\'e Grenoble Alpes and the hospitality of the LPSC Grenoble for a research visit to Grenoble at the final stage of this work. 
H.R.G.~also acknowledges the hospitality and financial support of the LPSC Grenoble trough the BSMGA project. 


\appendix

\clearpage 
\section{Overview of Run~2 results in the \smo\ 2.1.0 database}
\label{app:analysistables}

Tables~\ref{table:atlas} and \ref{table:cms} present a list of all Run 2 results included in the \smo\ database. The last column in each table displays which type of information is available (if any) for combining distinct signal regions within a given analysis.

\begin{table}[h!]
\begin{tabular}{|l|l|c|c|c|c|c|c|c|}
\hline
{\bf ID} & {\bf Short Description} & {\bf $\mathcal{L}$ [fb$^{-1}$] } & {\bf UL$_\mathrm{obs}$} & {\bf UL$_\mathrm{exp}$} & {\bf EM}& {\bf comb.}\\
\hline
\href{https://atlas.web.cern.ch/Atlas/GROUPS/PHYSICS/PAPERS/SUSY-2015-01/}{ATLAS-SUSY-2015-01}~\cite{Aaboud:2016nwl} & 2 $b$-jets  &3.2 & \checkmark &   &  &   \\
\href{https://atlas.web.cern.ch/Atlas/GROUPS/PHYSICS/PAPERS/SUSY-2015-02/}{ATLAS-SUSY-2015-02}~\cite{Aaboud:2016lwz} & $1\ell$ stop &3.2 & \checkmark &   & \checkmark&   \\
\href{http://atlas.web.cern.ch/Atlas/GROUPS/PHYSICS/PAPERS/SUSY-2015-06/}{ATLAS-SUSY-2015-06}~\cite{Aaboud:2016zdn} & $0\ell$ + 2--6 jets  &3.2 &   &   & \checkmark&   \\
\href{https://atlas.web.cern.ch/Atlas/GROUPS/PHYSICS/PAPERS/SUSY-2015-09/}{ATLAS-SUSY-2015-09}~\cite{Aad:2016tuk} & jets + 2 SS or $\ge 3\ell$ & 3.2 & \checkmark &   &  &   \\
\href{https://atlas.web.cern.ch/Atlas/GROUPS/PHYSICS/PAPERS/SUSY-2016-06/}{ATLAS-SUSY-2016-06}~\cite{Aaboud:2017mpt} & disappearing tracks &36.1 &   &   & \checkmark&   \\
\href{https://atlas.web.cern.ch/Atlas/GROUPS/PHYSICS/PAPERS/SUSY-2016-07/}{ATLAS-SUSY-2016-07}~\cite{Aaboud:2017vwy} & $0\ell$ + jets  &36.1 & \checkmark &   & \checkmark&   \\
\href{https://atlas.web.cern.ch/Atlas/GROUPS/PHYSICS/PAPERS/SUSY-2016-08/}{ATLAS-SUSY-2016-08}~\cite{Aaboud:2017iio} & displaced vertices &32.8 & \checkmark &   &  &   \\
\href{http://atlas.web.cern.ch/Atlas/GROUPS/PHYSICS/PAPERS/SUSY-2016-14/}{ATLAS-SUSY-2016-14}~\cite{Aaboud:2017dmy} & 2 SS or 3 $\ell$'s + jets  &36.1 & \checkmark &   &  &   \\
\href{https://atlas.web.cern.ch/Atlas/GROUPS/PHYSICS/PAPERS/SUSY-2016-15/}{ATLAS-SUSY-2016-15}~\cite{Aaboud:2017ayj} & 0$\ell$ stop &36.1 & \checkmark &   &  &   \\
\href{https://atlas.web.cern.ch/Atlas/GROUPS/PHYSICS/PAPERS/SUSY-2016-16/}{ATLAS-SUSY-2016-16}~\cite{Aaboud:2017aeu} & 1$\ell$ stop &36.1 & \checkmark &   & \checkmark&   \\
\href{http://atlas.web.cern.ch/Atlas/GROUPS/PHYSICS/PAPERS/SUSY-2016-17/}{ATLAS-SUSY-2016-17}~\cite{Aaboud:2017nfd} & 2 OS leptons  &36.1 & \checkmark &   &  &   \\
\href{https://atlas.web.cern.ch/Atlas/GROUPS/PHYSICS/PAPERS/SUSY-2016-19/}{ATLAS-SUSY-2016-19}~\cite{Aaboud:2018kya} & 2 $b$-jets + $\tau$'s 
&36.1 & \checkmark &   &  &   \\
\href{https://atlas.web.cern.ch/Atlas/GROUPS/PHYSICS/PAPERS/SUSY-2016-24/}{ATLAS-SUSY-2016-24}~\cite{Aaboud:2018jiw} & 2--3 $\ell$'s, EWino &36.1 & \checkmark &   & \checkmark&   \\
\href{https://atlas.web.cern.ch/Atlas/GROUPS/PHYSICS/PAPERS/SUSY-2016-26/}{ATLAS-SUSY-2016-26}~\cite{Aaboud:2018zjf} & $\ge 2\,c$-jets  &36.1 & \checkmark &   &  &   \\
\href{https://atlas.web.cern.ch/Atlas/GROUPS/PHYSICS/PAPERS/SUSY-2016-27/}{ATLAS-SUSY-2016-27}~\cite{Aaboud:2018doq} & jets + $\gamma$  &36.1 & \checkmark &   & \checkmark&   \\
\href{https://atlas.web.cern.ch/Atlas/GROUPS/PHYSICS/PAPERS/SUSY-2016-28/}{ATLAS-SUSY-2016-28}~\cite{Aaboud:2017wqg} & 2 $b$-jets  &36.1 & \checkmark &   &  &   \\
\href{http://atlas.web.cern.ch/Atlas/GROUPS/PHYSICS/PAPERS/SUSY-2016-32/index.html}{ATLAS-SUSY-2016-32}~\cite{Aaboud:2019trc} & HSCP &31.6 & \checkmark & \checkmark & \checkmark&   \\
\href{https://atlas.web.cern.ch/Atlas/GROUPS/PHYSICS/PAPERS/SUSY-2016-33/}{ATLAS-SUSY-2016-33}~\cite{Aaboud:2018ujj} & 2 OSSF $\ell$'s  &36.1 & \checkmark &   &  &   \\
\href{https://atlas.web.cern.ch/Atlas/GROUPS/PHYSICS/PAPERS/SUSY-2017-01/}{ATLAS-SUSY-2017-01}~\cite{Aaboud:2018ngk} & $WH(bb)$, EWino &36.1 & \checkmark &   &  &   \\
\href{https://atlas.web.cern.ch/Atlas/GROUPS/PHYSICS/PAPERS/SUSY-2017-02/}{ATLAS-SUSY-2017-02}~\cite{Aaboud:2018htj} & 0$\ell$ + jets  &36.1 & \checkmark & \checkmark &  &   \\
\href{https://atlas.web.cern.ch/Atlas/GROUPS/PHYSICS/PAPERS/SUSY-2017-03/}{ATLAS-SUSY-2017-03}~\cite{Aaboud:2018sua} & multi-$\ell$ EWino &36.1 & \checkmark &   & \checkmark&   \\
\href{https://atlas.web.cern.ch/Atlas/GROUPS/PHYSICS/PAPERS/SUSY-2018-04/}{ATLAS-SUSY-2018-04}~\cite{Aad:2019byo} & 2 hadronic taus &139.0 & \checkmark &   & \checkmark& JSON \\
\href{https://atlas.web.cern.ch/Atlas/GROUPS/PHYSICS/PAPERS/SUSY-2018-06/}{ATLAS-SUSY-2018-06}~\cite{Aad:2019vvi} & 3 leptons, EWino &139.0 & \checkmark & \checkmark & \checkmark&   \\
\href{https://atlas.web.cern.ch/Atlas/GROUPS/PHYSICS/PAPERS/SUSY-2018-10/}{ATLAS-SUSY-2018-10}~\cite{ATLAS:2021twp} & 1$\ell$ + jets  &139.0 & \checkmark &   & \checkmark&   \\
\href{https://atlas.web.cern.ch/Atlas/GROUPS/PHYSICS/PAPERS/SUSY-2018-12/}{ATLAS-SUSY-2018-12}~\cite{ATLAS:2020dsf} & $0\ell$ + jets  &139.0 & \checkmark & \checkmark & \checkmark&   \\
\href{https://atlas.web.cern.ch/Atlas/GROUPS/PHYSICS/PAPERS/SUSY-2018-14/}{ATLAS-SUSY-2018-14}~\cite{ATLAS:2020wjh} & displaced leptons &139.0 &   &   & \checkmark& JSON \\
\href{https://atlas.web.cern.ch/Atlas/GROUPS/PHYSICS/PAPERS/SUSY-2018-22/}{ATLAS-SUSY-2018-22}~\cite{ATLAS:2020syg} & multi-jets  &139.0 & \checkmark &   & \checkmark&   \\
\href{https://atlas.web.cern.ch/Atlas/GROUPS/PHYSICS/PAPERS/SUSY-2018-23/}{ATLAS-SUSY-2018-23}~\cite{ATLAS:2020qlk} & $WH(\gamma\gamma)$, EWino &139.0 & \checkmark & \checkmark &  &   \\
\href{https://atlas.web.cern.ch/Atlas/GROUPS/PHYSICS/PAPERS/SUSY-2018-31/}{ATLAS-SUSY-2018-31}~\cite{Aad:2019pfy} & 2$b$ + 2$H(bb)$  &139.0 & \checkmark &   & \checkmark& JSON \\
\href{https://atlas.web.cern.ch/Atlas/GROUPS/PHYSICS/PAPERS/SUSY-2018-32/}{ATLAS-SUSY-2018-32}~\cite{Aad:2019vnb} & 2 OS leptons  &139.0 & \checkmark &   &  &   \\
\href{https://atlas.web.cern.ch/Atlas/GROUPS/PHYSICS/PAPERS/SUSY-2019-08/}{ATLAS-SUSY-2019-08}~\cite{Aad:2019vvf} & 1$\ell$ + $H(bb)$, EWino &139.0 & \checkmark &   & \checkmark& JSON \\
\hline
\end{tabular}
\caption{List of the 31 ATLAS Run~2 analyses and their types of results in the \smo\ 2.1.0 database. Apart from the HSCP, DT and displaced lepton searches, all analyses require $\etmiss$ in the final state (for conciseness omitted in the short descriptions).\label{table:atlas}}
\end{table}\clearpage

\begin{table}[h!]
\begin{tabular}{|l|l|c|c|c|c|c|c|c|}
\hline
{\bf ID} & {\bf Short Description} & {\bf $\mathcal{L}$ [fb$^{-1}$] } & {\bf UL$_\mathrm{obs}$} & {\bf UL$_\mathrm{exp}$} & {\bf EM}& {\bf comb.}\\
\hline
\href{http://cms-results.web.cern.ch/cms-results/public-results/preliminary-results/EXO-16-036/index.html}{CMS-PAS-EXO-16-036}~\cite{CMS-PAS-EXO-16-036} & HSCP &12.9 & \checkmark &   &  &   \\
\href{http://cms-results.web.cern.ch/cms-results/public-results/preliminary-results/SUS-16-052/index.html}{CMS-PAS-SUS-16-052}~\cite{CMS-PAS-SUS-16-052} & ISR jet + soft $\ell$ &35.9 & \checkmark &   & \checkmark& Cov. \\
\href{https://cms-results.web.cern.ch/cms-results/public-results/publications/SUS-16-009/}{CMS-SUS-16-009}~\cite{Khachatryan:2017rhw} & $0\ell$ + jets, top tag &2.3 & \checkmark & \checkmark &  &   \\
\href{http://cms-results.web.cern.ch/cms-results/public-results/publications/SUS-16-032/index.html}{CMS-SUS-16-032}~\cite{Sirunyan:2017kiw} & 2 $b$- or 2 $c$-jets &35.9 & \checkmark &   &  &   \\
\href{http://cms-results.web.cern.ch/cms-results/public-results/publications/SUS-16-033/index.html}{CMS-SUS-16-033}~\cite{Sirunyan:2017cwe} & 0$\ell$ + jets  &35.9 & \checkmark & \checkmark & \checkmark&   \\
\href{http://cms-results.web.cern.ch/cms-results/public-results/publications/SUS-16-034/index.html}{CMS-SUS-16-034}~\cite{Sirunyan:2017qaj} & 2 OSSF leptons &35.9 & \checkmark &   &  &   \\
\href{http://cms-results.web.cern.ch/cms-results/public-results/publications/SUS-16-035/index.html}{CMS-SUS-16-035}~\cite{Sirunyan:2017uyt} & 2 SS leptons &35.9 & \checkmark &   &  &   \\
\href{http://cms-results.web.cern.ch/cms-results/public-results/publications/SUS-16-036/index.html}{CMS-SUS-16-036}~\cite{Sirunyan:2017kqq} & 0$\ell$ + jets  &35.9 & \checkmark & \checkmark &  &   \\
\href{http://cms-results.web.cern.ch/cms-results/public-results/publications/SUS-16-037/index.html}{CMS-SUS-16-037}~\cite{Sirunyan:2017fsj} & 1$\ell$ + jets  with MJ &35.9 & \checkmark &   &  &   \\
\href{http://cms-results.web.cern.ch/cms-results/public-results/publications/SUS-16-039/index.html}{CMS-SUS-16-039}~\cite{Sirunyan:2017lae} & multi-$\ell$, EWino &35.9 & \checkmark &   &  &   \\
\href{http://cms-results.web.cern.ch/cms-results/public-results/publications/SUS-16-041/index.html}{CMS-SUS-16-041}~\cite{Sirunyan:2017hvp} & multi-$\ell$ + jets  &35.9 & \checkmark &   &  &   \\
\href{http://cms-results.web.cern.ch/cms-results/public-results/publications/SUS-16-042/index.html}{CMS-SUS-16-042}~\cite{Sirunyan:2017mrs} & 1$\ell$ + jets  &35.9 & \checkmark &   &  &   \\
\href{http://cms-results.web.cern.ch/cms-results/public-results/publications/SUS-16-043/index.html}{CMS-SUS-16-043}~\cite{Sirunyan:2017zss} & $WH(bb)$, EWino &35.9 & \checkmark &   &  &   \\
\href{http://cms-results.web.cern.ch/cms-results/public-results/publications/SUS-16-045/index.html}{CMS-SUS-16-045}~\cite{Sirunyan:2017eie} & 2 $b$ + 2 $H(\gamma\gamma)$ &35.9 & \checkmark &   &  &   \\
\href{http://cms-results.web.cern.ch/cms-results/public-results/publications/SUS-16-046/index.html}{CMS-SUS-16-046}~\cite{Sirunyan:2017nyt} & high-$p_T$ $\gamma$  &35.9 & \checkmark &   &  &   \\
\href{http://cms-results.web.cern.ch/cms-results/public-results/publications/SUS-16-047/index.html}{CMS-SUS-16-047}~\cite{Sirunyan:2017yse} & $\gamma$ + jets, high $H_T$ &35.9 & \checkmark &   &  &   \\
\href{http://cms-results.web.cern.ch/cms-results/public-results/publications/SUS-16-049/index.html}{CMS-SUS-16-049}~\cite{Sirunyan:2017wif} & 0$\ell$ stop &35.9 & \checkmark & \checkmark &  &   \\
\href{http://cms-results.web.cern.ch/cms-results/public-results/publications/SUS-16-050/index.html}{CMS-SUS-16-050}~\cite{Sirunyan:2017pjw} & 0$\ell$ + top tag &35.9 & \checkmark & \checkmark &  &   \\
\href{http://cms-results.web.cern.ch/cms-results/public-results/publications/SUS-16-051/index.html}{CMS-SUS-16-051}~\cite{Sirunyan:2017xse} & 1$\ell$ stop &35.9 & \checkmark & \checkmark &  &   \\
\href{http://cms-results.web.cern.ch/cms-results/public-results/publications/SUS-17-001/index.html}{CMS-SUS-17-001}~\cite{Sirunyan:2017leh} & 2$\ell$ stop &35.9 & \checkmark &   &  &   \\
\href{https://cms-results.web.cern.ch/cms-results/public-results/publications/SUS-17-003/}{CMS-SUS-17-003}~\cite{Sirunyan:2018vig} & 2 taus  &35.9 & \checkmark &   &  &   \\
\href{http://cms-results.web.cern.ch/cms-results/public-results/publications/SUS-17-004/index.html}{CMS-SUS-17-004}~\cite{Sirunyan:2018ubx} & EWino combination &35.9 & \checkmark &   &  &   \\
\href{https://cms-results.web.cern.ch/cms-results/public-results/publications/SUS-17-005/}{CMS-SUS-17-005}~\cite{Sirunyan:2018omt} & 1$\ell$ + jets, top tag &35.9 & \checkmark & \checkmark &  &   \\
\href{https://cms-results.web.cern.ch/cms-results/public-results/publications/SUS-17-006/}{CMS-SUS-17-006}~\cite{Sirunyan:2017bsh} & jets + boosted $H(bb)$  &35.9 & \checkmark & \checkmark &  &   \\
\href{https://cms-results.web.cern.ch/cms-results/public-results/publications/SUS-17-009/}{CMS-SUS-17-009}~\cite{Sirunyan:2018nwe} & SFOS leptons  &35.9 & \checkmark & \checkmark &  &   \\
\href{http://cms-results.web.cern.ch/cms-results/public-results/publications/SUS-17-010}{CMS-SUS-17-010}~\cite{Sirunyan:2018lul} & 2$\ell$ stop &35.9 & \checkmark & \checkmark &  &   \\
\href{https://cms-results.web.cern.ch/cms-results/public-results/publications/SUS-18-002/}{CMS-SUS-18-002}~\cite{Sirunyan:2019hzr} & $\gamma$ +  ($b$-)jets, top tag &35.9 & \checkmark & \checkmark &  &   \\
\href{http://cms-results.web.cern.ch/cms-results/public-results/publications/SUS-19-006/index.html}{CMS-SUS-19-006}~\cite{Sirunyan:2019ctn} & 0$\ell$ + jets, MHT &137.0 & \checkmark & \checkmark &  &   \\
\href{http://cms-results.web.cern.ch/cms-results/public-results/publications/SUS-19-009/index.html}{CMS-SUS-19-009}~\cite{Sirunyan:2019glc} & 1$\ell$ + jets, MHT &137.0 & \checkmark &   &  &   \\
\href{http://cms-results.web.cern.ch/cms-results/public-results/publications/EXO-19-001/index.html}{CMS-EXO-19-001}~\cite{Sirunyan:2019gut} & non-prompt jets &137.0 &   &   & \checkmark&   \\
\href{http://cms-results.web.cern.ch/cms-results/public-results/publications/EXO-19-010/}{CMS-EXO-19-010}~\cite{CMS:2020atg} & disappearing tracks &101.0 &   &   & \checkmark&   \\
\hline
\end{tabular}
\caption{List of the 31 CMS Run~2 analyses and their types of results in the \smo\ 2.1.0 database. In the last column, ``Cov.'' stands for covariance matrix. 
All CMS-SUS analyses require $\etmiss$ in the final state (for conciseness omitted in the short descriptions).\label{table:cms}}
\end{table}
 
\clearpage 
\section{Recasting of the ATLAS-SUSY-2016-32 HSCP search} \label{app:RecastHSCP}

For the recasting of the 13~TeV ATLAS search for HSCPs~\cite{Aaboud:2019trc}, we follow the prescription provided in the auxiliary information of the publication. 
We consider the two signal regions
\texttt{SR-1Cand-FullDet} and \texttt{SR-2Cand-FullDet} that 
use the time-of-flight measurement for reconstructing the mass of the HSCP\@. The signal region \texttt{SR-1Cand-FullDet} requires one (and only one) HSCP candidate that passes the `tight' selection criterion while the signal region \texttt{SR-2Cand-FullDet} requires candidates satisfying the `loose' selection criteria. In the analysis, two different triggers are considered, an $\etmiss$ trigger and a muon trigger. 
Since we consider only constraints from HSCPs decaying outside the muon chamber, the recasting assumes the muon trigger.

The probabilities for an HSCP candidate to pass the muon trigger and satisfy the loose and tight selection criteria are given as a function of its velocity, $\beta$, and pseudorapidity, $\eta$, at generator level in the auxiliary information of~\cite{Aaboud:2019trc}. We denote them by $P_\text{trig}$, $P_\text{loose}$ and $P_\text{tight}$, respectively. (Note that $P_\text{tight} = P_\text{loose} \,P_\text{tight-promotion}$.) Furthermore, for each of the two signal regions, the analysis considers four different choices for the cut on the reconstructed mass, $m_\text{reco}$. 
This gives a total of eight kinematic regions for which EMs are needed for the \smo\ database.

We compute the probability for the reconstructed mass to lie above the respective cut by assuming $m_\text{reco}$ to be Gaussian distributed. The respective mean value and variance has been provided as a function of the true mass in the auxiliary information of~\cite{Aaboud:2019trc}. We denote this probability with $P_{m_\text{cut}}$.
We furthermore have to consider the probability that the given HSCP in an event traverses the full detector, $F_\text{long}$. The latter is given by
\begin{equation}
F_\text{long}(\beta,\eta)=\exp\,\left(-\frac{L(\eta)}{c\tau \gamma\beta}\right)
\end{equation}
where $\gamma$ is the relativistic boost factor according to the velocity $\beta$ and $L(\eta)$ is the $\eta$-dependent travel length of the HSCP traversing the detector. To compute $L(\eta)$, we approximate the ATLAS detector by a cylinder with a radius of 12~m and a length of 46~m.

For a given event with two HSCP candidates (that have the same mass), we compute the overall probability that the event is triggered and selected in a given signal region with $m_\text{reco}$ above the respective cut by
\begin{equation}
\begin{split}
    {\cal P}^\text{event}_{\text{1Cand},m_\text{cut}} = &\;\Big\{
        F_\text{long}^{(1)} \left( 1 - F_\text{long}^{(2)}\right)
        P_\text{trig}^{(1)} P_\text{tight}^{(1)} 
         + F_\text{long}^{(2)} \left( 1 - F_\text{long}^{(1)}\right)
        P_\text{trig}^{(2)} P_\text{tight}^{(2)} \\
        &   \quad +F_\text{long}^{(1)} F_\text{long}^{(2)} \left(
            P_\text{trig}^{(1)}+ P_\text{trig}^{(2)}-
            P_\text{trig}^{(1)} P_\text{trig}^{(2)}\right)\\
        & \qquad\times \left[ P_\text{tight}^{(1)} \left( 1 - P_\text{loose}^{(2)} \right)
         + P_\text{tight}^{(2)} \left( 1 - P_\text{loose}^{(1)} \right)\right]\Big\} P_{m_\text{cut}}
\end{split}
\end{equation}
for \texttt{SR-1Cand-FullDet}, and 
\begin{equation}
    {\cal P}^\text{event}_{\text{2Cand},m_\text{cut}} = 
    \,F_\text{long}^{(1)} \,F_\text{long}^{(2)} P_\text{loose}^{(1)}  P_\text{loose}^{(2)}
    \times \left( P_\text{trig}^{(1)} + P_\text{trig}^{(2)} - P_\text{trig}^{(1)}  P_\text{trig}^{(2)} \right) P_{m_\text{cut}}^2
\end{equation}
for \texttt{SR-2Cand-FullDet}. 
From the above equations, if one HSCP particle is present in each event, 
only the \texttt{SR-1Cand-FullDet} is applicable and 
${\cal P}^\text{event}_{\text{1Cand},m_\text{cut}}=F_\text{long} 
        P_\text{trig} P_\text{tight} P_{m_\text{cut}} $.

We derive the EMs for the above-mentioned eight signal regions by generating events utilizing~\textsc{MadGraph5\_aMC@NLO}~\cite{Alwall:2014hca}
for the hard scattering, and \textsc{Pythia}~8~\cite{Sjostrand:2014zea} for showering and hadronization. We do not perform any detector simulation as the recasting is based on the kinematics of the HSCP at generator level. (In fact, hadronization only affects the isolation criterion requiring the sum of the track-$p_\text{T}$ in a cone of $\Delta R = 0.2$ around the candidate's track to be below 5~GeV. However, the effect of the isolation criterion on the EMs is small.)
For each simplified model parameter point in the EM, we generate $N=2.5\times 10^4$ events and compute the `efficiency', $({\cal A}\varepsilon)_{\text{SR}}$, for a given signal region (SR) by
\begin{equation}
    ({\cal A}\varepsilon)_{\text{SR}} = \frac{1}{N}\sum_{i=1}^N {\cal P}^{i}_{\text{SR}}\,.
\end{equation}

\begin{figure}[t!]\centering
 \includegraphics[width=0.8\textwidth]{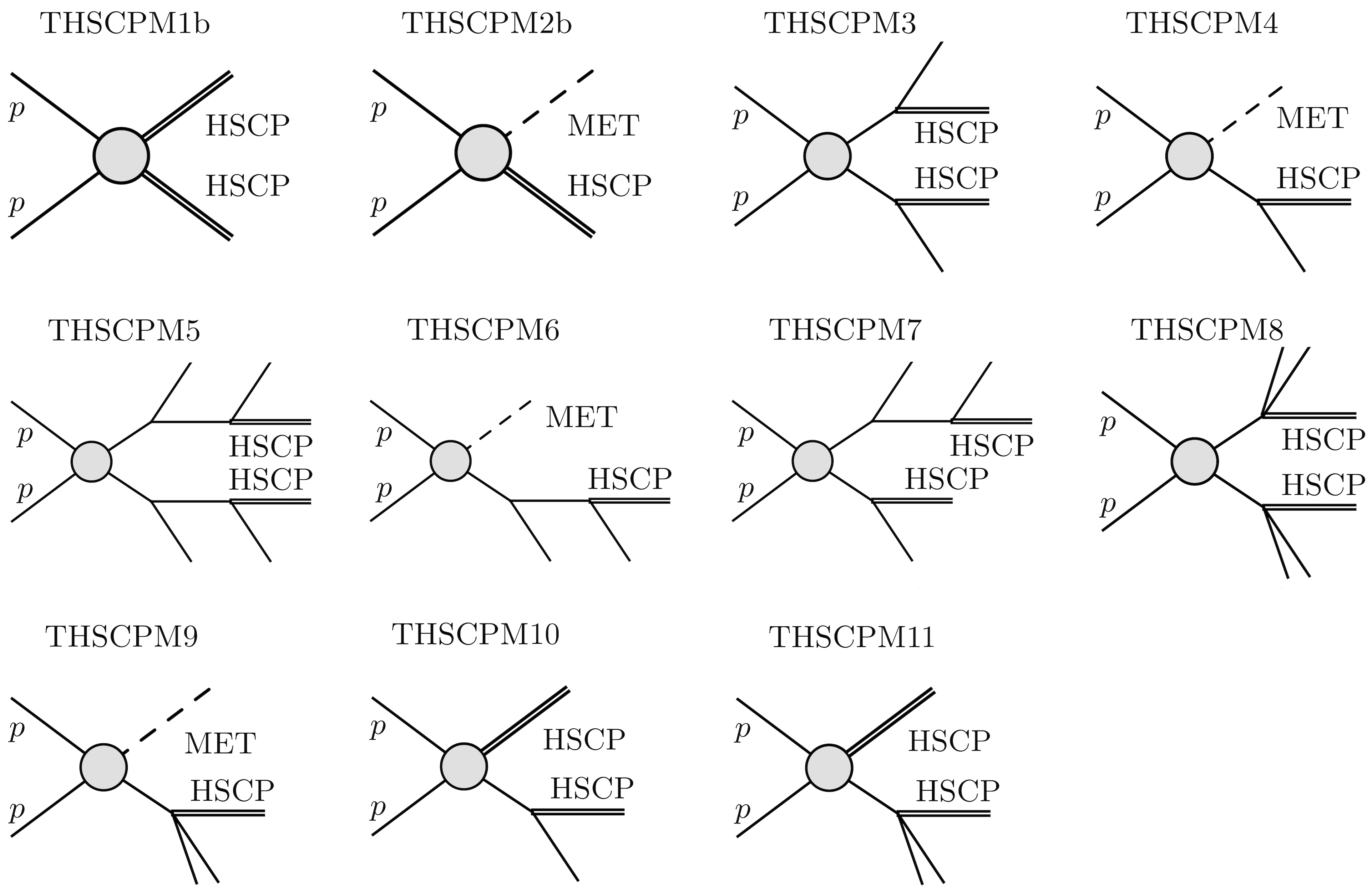}
 \caption{ \label{fig:THSCPMdiag} 
 Illustration of the 11 topologies for which we have generated EMs for the ATLAS-SUSY-2016-32 HSCP search.
}
\end{figure}

We perform the computation of EMs for the 11 topologies depicted in Fig.~\ref{fig:THSCPMdiag}. In four of the topologies (THSCPM2b, THSCPM4, THSCPM6, and THSCPM9) only one branch contains an HSCP, while the other branch is assumed to terminate in a neutral particle. All other topologies have two HSCP candidates. 
As the search relies only on the HSCP itself, it is largely\footnote{The isolation criterion, in principle, introduces a dependence on the type of SM particle in the event. However, the reduction of efficiencies due to the isolation criterion is small, in particular, compared to expected uncertainties introduced by the approximations associated with the simplified model assumptions.} insensitive to the type of SM particles being emitted in the decays: 
the dependence on the type of topology and BSM masses involved only enters through its effect on the kinematics of the HSCP\@. Similarly, for the four topologies with only one HSCP, the EMs are insensitive to the \etmiss (MET) branch except for the mass of the parent particle produced by the hard scattering as it affects the kinematics of the HSCP branch. Accordingly, we only specify the first particle of the MET branch indicated
by a dashed line in the respective diagrams in Fig.~\ref{fig:THSCPMdiag}; this representation is meant to
implicitly include all possible cascade decays.  

For all topologies that involve up to two mass parameters, we take into account the explicit width dependence in the database. For the three topologies which involve a 2-step cascade decay (THSCPM5, THSCPM6, and THSCPM7) and, hence, three mass parameters, we employ the detector-stable limit only, keeping the EM grids to be three-dimensional at most. This is done to limit the size of the database pickle file.

The simplified model parameters are summarized in Table~\ref{tab:HSCPtopo}. For the computation of the EMs, we use realizations of the topologies within the MSSM\@. The respective processes are also included in Table~\ref{tab:HSCPtopo}. Note that for the four topologies with one HSCP only, we re-use the events of the respective process involving two HSCPs in our analysis by taking into account only one of the candidates at a time.

\begin{table}[t]
\begin{center}
\renewcommand{\arraystretch}{1.5}
\begin{tabular}{|c  c  c |} 
\hline
  Topology name & free parameters & SUSY process \\
 \hline
THSCPM1b & $m_\text{HSCP},\Gamma_\text{HSCP}$ & $p p\to \tilde\tau \tilde \tau $\\
THSCPM2b & $m_\text{HSCP},\Gamma_\text{HSCP}$ & $p p\to \tilde\tau \tilde \tau^* $\\
THSCPM3 & $m_\text{prod},m_\text{HSCP},\Gamma_\text{HSCP}$ & $p p\to \tilde q \tilde q \to \tilde \chi^\pm \tilde \chi^\pm $\\
THSCPM4 & $m_\text{prod},m_\text{HSCP},\Gamma_\text{HSCP}$ & $p p\to \tilde q \tilde q \to \tilde \chi^\pm \tilde \chi^\pm{}^* $\\
THSCPM5 & $m_\text{prod},m_\text{int},m_\text{HSCP}$ & $p p\to \tilde q \tilde q \to \tilde \chi^0 \tilde \chi^0 \to \tilde\tau \tilde \tau$\\
THSCPM6 & $m_\text{prod},m_\text{int},m_\text{HSCP}$ & $p p\to \tilde q \tilde q \to \tilde \chi^0 \tilde \chi^0 \to \tilde\tau \tilde \tau^*$\\
THSCPM7 & $m_\text{prod},m_\text{int},m_\text{HSCP}$ & $p p\to \tilde \chi^0 \tilde \chi^\pm_2 \to \tilde \tau (\tilde \chi^\pm_1 \to \tilde \tau)$\\
THSCPM8 & $m_\text{prod},m_\text{HSCP},\Gamma_\text{HSCP}$ & $p p\to \tilde q \tilde q \to \tilde\tau \tilde \tau $\\
THSCPM9 & $m_\text{prod},m_\text{HSCP},\Gamma_\text{HSCP}$ & $p p\to \tilde q \tilde q \to \tilde\tau \tilde \tau^* $\\
THSCPM10 & $m_\text{prod},m_\text{HSCP},\Gamma_\text{HSCP}$ & $p p\to \tilde \chi^\pm ( \tilde q \to \tilde \chi^\pm) $\\
THSCPM11 & $m_\text{prod},m_\text{HSCP},\Gamma_\text{HSCP}$ & $p p\to \tilde \chi^\pm ( \tilde g \to \tilde \chi^\pm) $\\
\hline
\end{tabular}
\renewcommand{\arraystretch}{1}
\end{center}
\caption{Simplified model parameters and the SUSY processes used in the simulation of the 11 topologies shown in Fig.~\ref{fig:THSCPMdiag}. 
$^*$\,Only one of the HSCP candidates is taken into account.
}
\label{tab:HSCPtopo}
\end{table}

\clearpage
\section{Updating the \micromegasversion\ interface}
\label{app:MOinterface}

For using \smo\,v2.1 with  \micromegasversion, the interface files

\verb|micromegas_5.2.7.a/include/SMODELS.inc|, 

\verb|micromegas_5.2.7.a/sources/smodels.c|, and 

\verb|micromegas_5.2.7.a/Packages/SMODELS.makef|  

\noindent
should be replaced by the ones provided with this paper \cite{MO_SMO_zenodo}. 
We also recommend to update the parameters file 
\verb|micromegas_5.2.7.a/include/smodels_parameters.ini| 
with the one provided here. 

In the \micromegas\ main program, \smo\ can then be called with 
the code snippet below (also included in the example main program on \cite{MO_SMO_zenodo}).  

\begin{verbatim}
#ifdef SMODELS
{ int status=0, smodelsOK=0; 
  double Rvalue, Rexpected, SmoLsig, SmoLmax, SmoLSM;
  char analysis[50]={},topology[100]={},smodelsInfo[100];
  int LHCrun=LHC8|LHC13;  //  LHC8  - 8TeV; LHC13  - 13TeV;   

  printf("\n\n=====  LHC constraints with SModelS  =====\n\n");

#include "../include/SMODELS.inc" // SLHA interface with SModelS

  printf("SModelS %s \n",smodelsInfo);
  if(smodelsOK) 
  { printf(" highest r-value = %.2E",Rvalue); 
  
    if(Rvalue>0) 
    { printf(" from %s, topology: %s ",analysis,topology);
      if(Rexpected>0) 
      { printf("\n expected r = %.2E ",Rexpected);
        if(SmoLsig>0) 
        { printf("\n -2log (L_signal, L_max, L_SM) = %.2E %.2E %.2E", 
                  -2*log(SmoLsig),-2*log(SmoLmax),-2*log(SmoLSM)); }
      }
    }  
    if(status==1) printf("\n excluded by SMS results"); 
    else if(status==0) printf("\n not excluded"); 
    else if(status==-1) printf("\n not not tested by results in SModelS database"); 
    printf("\n");
  } else system("cat smodels.err"); // problem: see smodels.err
}   
#endif 
\end{verbatim}

For the \verb|mssms.par| parameter point in the MSSM directory of \micromegas, for instance, this will give 

\begin{verbatim}
=====  LHC constraints with SModelS  =====

found SM-like Higgs = h
writing mass block and decay tables ... 
computing LHC cross sections ... 
SLHA input file done.

SModelS v2.1.1 with database 2.1.0 
 highest r-value = 1.07E+01 from ATLAS-SUSY-2018-22, topology: T2  
 excluded by SMS results
\end{verbatim}

If \verb|#define CLEAN| is commented out in \micromegas, the input and output files for \smo\ (\ie\ the \verb|smodels.slha| file containing the mass spectrum, decay tables and cross sections, and the  \verb|smodels.slha.smodelsslha| file containing the \smo\ results) will be kept. 
With standard settings, \verb|smodels.slha.smodelsslha| reports all the excluding experimental results for excluded points but only the most constraining result for non-excluded points; to always have all applicable results listed, set 
\begin{verbatim}
[slha-printer]
expandedOutput = True 
\end{verbatim}
in the \verb|smodels_parameters.ini| file. 
Other options, like \verb|testCoverage|, 
\verb|combineSRs|, etc., can also be turned on/off via the  
\verb|smodels_parameters.ini| file. 
Other output formats can be chosen via the 
\begin{verbatim}
[printer]
outputType = ...
\end{verbatim}
option. Available formats are \verb|slha|, \verb|summary|, \verb|python|, \verb|xml| and \verb|stdout|; note however that \verb|slha| output is always necessary for \micromegas.

The updated interface to \smo\,v2.1 (or higher) is included by default from \micromegas\,v5.2.10  onward. Note that the exact \smo\ version to be downloaded and included in \micromegas\  \verb|Packages/| is set in the \verb|include/SMODELS.inc| file.
Additional information is given in \cite{Barducci:2016pcb} and the manual shipped with the \micromegas\ distribution.


\clearpage
\providecommand{\href}[2]{#2}\begingroup\raggedright\endgroup

\end{document}